\documentclass[%
aip,
jcp,%
amsmath,amssymb,
reprint,%
longbibliography
]{revtex4-1}
\usepackage[latin1]{inputenc}

\usepackage{amsmath}%
\usepackage{amsfonts}
\usepackage{amssymb,bm}
\usepackage{graphicx}
\usepackage{physics}
\usepackage{upgreek}
\usepackage[outercaption]{sidecap}   
\usepackage{color}
\usepackage[normalem]{ulem}
\mathchardef\mhyphen="2D
\usepackage{txfonts}

\usepackage[nice]{nicefrac}

\DeclareMathAlphabet{\pazocal}{OMS}{zplm}{m}{n}

\date{\today}

\begin{document}

\def\bra#1{\left<{#1}\right|}
\def\ket#1{\left|{#1}\right>}
\def\expval#1#2{\bra{#2} {#1} \ket{#2}}
\def\mapright#1{\smash{\mathop{\longrightarrow}\limits^{_{_{\phantom{X}}}{#1}_{_{\phantom{X}}}}}}

\title{A unified framework for semiclassical reaction rate theory}

\author{Joseph E.\ Lawrence}
\affiliation{\mbox{Simons Center for Computational Physical Chemistry, New York University, New York, NY 10003, USA}
\\
Department of Chemistry, New York University, New York, NY 10003, USA}
\email{joseph.lawrence@nyu.edu}

\begin{abstract}

A general semiclassical theory for the calculation of reaction rate constants is developed. 
The theory can be understood as a formal framework that encompasses existing semiclassical methods: instanton theory and semiclassical transition state theory (SCTST).
Unlike SCTST, the present formalism does not start from the concept of  ``good'' action-angle variables. Instead, it is based on a conjectured connection between the cumulative reaction probability and the instanton contribution to the formally exact generalisation of Gutzwiller's formula for the trace of the Green's function. The formalism effectively generalises the ``imaginary free-energy'' formulation of instanton theory to microcanonical scattering rates and all orders in $\hbar$. In one dimension, explicit expressions are derived for the generalised reduced action up to $\mathcal{O}(\hbar^4)$ using exact WKB/quantum Hamilton-Jacobi theory. 
The connection between the present formalism and the standard second order vibrational perturbation theory (VPT2) version of SCTST is explored. 
It is also shown that the standard thermal instanton rate theory, as well as higher order (dividing surface independent) ``perturbative'' corrections can be straightforwardly derived from the framework. Above the crossover temperature, first-order corrections in $\hbar$ to the parabolic barrier (``sphaleron'') rate are also derived. 
A simple anharmonic transition state theory and anharmonic version of the Wigner tunneling correction are presented. 
Finally, the potential for the development of new and improved semiclassical methods for modelling reaction kinetics is discussed.

\end{abstract}

\maketitle

\section{Introduction}
The progress in efficient and accurate electronic structure calculations,\cite{Sherrill2020JCPElectronicStructureIntro,Zheng2007BarrierHeights,Prasad2022BarrierHeights,Spiekermann2022BarrierHeights} automatic reaction network exploration,\cite{Chang2023RxnDiscoveryTodd,Segler2018RxnDiscoveryNN,Maeda2023RxnDiscovery} and master equation techniques\cite{Glowacki2012MESMER,Georgievskii2013MESS,Zhang2022TUMME,MultiWell} is opening up the possibility of fully ab initio reaction rate prediction. 
To realise this potential requires equally efficient and accurate methods for the calculation of elementary reaction rate constants. 
In order to integrate with automated protocols, methods for rate estimation need to be as black-box as possible, while still being highly accurate.
Ideally, these methods should retain the simple structure of rigid-rotor harmonic oscillator transition state theory (RRHO-TST), but be systematically improvable towards the fully exact result. %

Semiclassical analysis provides a natural starting point for the development of such a systematically improvable hierarchy of rate methods. Rigorous semiclassical analysis can be understood as a generalised form of perturbation theory, where the perturbative parameter is associated with the $\hbar$ of the phase factor, $e^{+iS/\hbar}$, that appears in the path-integral formulation of quantum mechanics.\cite{Tannor,Child,HellerBook} Hence, as with perturbation theory, one can in principle obtain systematic improvements to the theory by calculating higher order terms in the series.\cite{Kleinert,Lawrence2023RPI+PC} 
Semiclassical analysis has the additional benefit that the resulting expansion can be written entirely in terms of classical trajectories and their properties, providing an intuitive perspective on the quantum dynamics.
Rate constants of activated processes provide an ideal case for semiclassical techniques, as they are typically dominated by a single path near the transition state,\footnote{For a recent discussion of the challenges and methods for treating barrierless reactions see: C. Cavallotti, ``Automation of chemical kinetics: Status and challenges,'' Proc.  Combust. Inst. {\bf 39}, 11-28 (2023).} unlike other observables that require locating a large number of distinct classical trajectories.

The use of semiclassical methods to predict reaction rate constants has a long and rich history.\cite{Miller1998review,Truhlar2020SemiclassicalRateHistory} %
Here we focus in particular on two methods: instanton theory\cite{Miller1975semiclassical,Richardson2016FirstPrinciples,InstReview,Coleman1977ImF,Coleman1979UsesOfInstantons,Affleck1981ImF,Lawrence2024crossover} and semiclassical transition state theory (SCTST).\cite{Miller1977SCTST,Miller1990SCTST,Hernandez1993SCTST,Wagner2013SCTST,Shan2019SCTSTReview}
Crucially, both theories provide a practical extension to RRHO-TST, requiring only limited information in the vicinity of the transition state.\cite{Miller1990SCTST,GPR,Fang2024CheapInstantons} Instanton theory provides an accurate description of thermal tunneling in terms of the dominant tunneling path (the instanton) in a path-integral description of the rate, treating fluctuations harmonically, and can be practically applied to molecular systems\cite{GPR,Fang2024CheapInstantons,HCH4,Meisner2011isotope,Cooper2018interpolation,McConnell2019instanton,Litman2022InstantonElectronicFrictionI,Litman2022InstantonElectronicFrictionII,Zhang2025InstantonNEB} using the ``ring-polymer instanton'' (RPI) formulation.\cite{Andersson2009Hmethane,Richardson2009RPInst,InstReview} 
In contrast, SCTST gives an accurate description of anharmonicity about the transition state, but fails to describe ``deep'' tunneling. 
It is practically implemented using the machinery of vibrational perturbation theory,\cite{Miller1990SCTST} and has been widely applied to describe anharmonic effects in chemical reactions.\cite{Greene2016SCTST,Nguyen2010SCTST,Burd2018SCTSTTorsion,Mandelli2022SCTSTOrganicMolecules,Aieta2019SCTST,Conte2024PerspectiveSCTST_and_Spectra,Nguyen2025SCTST} %

Although both theories are based on semiclassical ideas, their historical derivations are qualitatively distinct. Instanton theory was originally derived in the chemical physics literature by Miller using Gutzwiller's periodic orbit theory.\cite{Miller1975semiclassical} More recently, it has been shown to be derivable from first principles using standard asymptotic (semiclassical) evaluation of the path-integral expression for the flux-correlation formulation of the  rate.\cite{Richardson2016FirstPrinciples,InstReview,Lawrence2024crossover} 
This has made possible the rigorous extension of the theory to electronically non-adiabatic systems,\cite{inverted,PhilTransA,thiophosgene,nitrene,4thorder,Fang2023ConicalIntersections,Ansari2024Oxygen,Richardson2024NonAdTunneling,Zarotiadis2025NASCI}  the inclusion of higher order perturbative corrections,\cite{Lawrence2023RPI+PC} and extension to arbitrary temperatures.\cite{Lawrence2024crossover}  
The derivation of SCTST, also due to Miller, combines the concept of ``good'' action-angle variables\cite{Miller1977SCTST} (an extension of the quantisation conditions of the old quantum theory to saddle points) with an analytic continuation of second order vibrational perturbation theory (VPT2).\cite{Miller1990SCTST}

While the 
qualitative similarities between the two theories are obvious, and recently suggestions have even been made to combine the methods,\cite{Upadhyayula2024hbar2corrections,Upadhyayula2025CollinearH+H2} their formal connection has remained largely unexplored. In the following we will build upon the ideas implicit in earlier work,\cite{Miller1977SCTST,Hernandez1993SCTST,Seideman1991SCTST_Siegert_eigvals} in combination with results from the modern mathematical physics literature,\cite{ChaosBook,Ture2025ExactWKB} to make the connection explicit.
Specifically, we will propose a unified semiclassical framework from which both theories can be derived.
This framework will not only clarify the approximations inherent in each method but also provide a basis for the derivation of a systematically improvable hierarchy of  methods that will be the subject of the following papers in this series.

Section \ref{sec:Motivation_Definitions} will introduce basic concepts of reaction rate theory used throughout the paper, and motivate the following section. Sec.~\ref{sec:General_Theory} will then introduce the central proposal of the paper, a unified semiclassical framework for reaction rate theory, based on the exact generalisation of Gutzwiller's trace formula. The conjecture of Sec.~\ref{sec:General_Theory} will be explored further in the one-dimensional case in Sec.~\ref{sec:One_dimension}. After this, Sec.~\ref{sec:SCTST} will explore the connection to VPT-SCTST. In Sec.~\ref{sec:instanton_connection} we will move to derive instanton theory from the new framework, as well as its first-order correction in $\hbar$.
Sec.~\ref{sec:sphaleron} will show how the instanton results generalise above the crossover temperature, giving the ``sphaleron'' rate and its first-order correction. 
Finally Sec.~\ref{sec:AnhTST_Wigner} will present a simple multidimensional anharmonic generalisation of the Wigner tunneling correction and corresponding anharmonic transition state theory, before Sec.~\ref{sec:Conclusion} concludes.

\section{Background and basic definitions} \label{sec:Motivation_Definitions}\renewcommand{\theequation}{2.\arabic{equation}}
\setcounter{equation}{0}
Our central focus in the present paper is the cumulative reaction probability, $N(E)$, which is related to the elementary thermal reaction rate constant, $k(\tau)$, according to
\begin{equation}
    k(\tau)Z_r(\tau) = \frac{1}{2\pi\hbar} \int_{-\infty}^\infty e^{-\tau E/\hbar} N(E) \mathrm{d}E, \label{eq:k_N_E_relation}
\end{equation}
where $Z_r(\tau)$ is the reactant partition function and $\tau=\beta\hbar=\hbar/(k_{\rm B}T)$ is the thermal time. In one dimension $N(E)$ is just the transmission probability, $P(E)$, for the barrier, and can be approximated semiclassically by the uniform WKB expression\cite{Kemble1935WKB,Froman1965JWKB} 
\begin{equation}
    P(E) \sim P_{\rm WKB}(E)= \frac{1}{1+e^{W_0(E)/\hbar}}   \label{eq:uniform_WKB_PE}
\end{equation}
as $\hbar\to0$, where 
\begin{equation}
    W_0(E) = 2\int_{x_{-}}^{x_+} \sqrt{2(V(x)-E)} \, \mathrm{d}x 
\end{equation}
is the classical (Euclidean) reduced action of the instanton orbit, and $x_{\pm}(E)$ are the mass-weighted turning points of the potential. 
Note that $A\sim B$ is read ``$A$ is asymptotically equal to $B$.'' The ``$\sim$'' symbol can be understood as representing the possibility that the perturbative series on the right hand side may be either, a truncated series, not formally convergent, or both. For a detailed discussion of asymptotic notation and asymptotic series see e.g.~Ref.~\citenum{BenderBook}.

In multiple dimensions, in the absence of rotations, the cumulative reaction probability can be written as a sum over the 
reactant vibrational states as 
\begin{equation}
    N(E)= \sum_{\mathbf{n}}P_{\mathbf{n},r}(E)
\end{equation}
where $P_{\mathbf{n},r}(E)$ can be interpreted as the probability that the system, initially in the reactant vibrational state described by the quantum numbers $\mathbf{n}=(n_1,n_2,\dots,n_{F-1})$
and incident on the barrier with a total energy $E$, will reach the products. 
Alternatively, by detailed balance one can 
 also resolve $N(E)$ in terms of a sum over the product vibrational states, 
 \begin{equation}
    N(E)= \sum_{\mathbf{n}}P_{\mathbf{n},p}(E)
\end{equation}
 where now the probabilities correspond to going from a given product state $\mathbf{n}$ at energy $E$ to any reactant state. 

To understand what follows it is sufficient to note two simple facts.
First, in principle one is not restricted to resolving $N(E)$ in a basis of states that correspond asymptotically to a specific reactant or product quantum state. 
Second, as is obvious from the fact that the above expressions hold for any given energy, one can choose the basis to resolve $N(E)$ to be energy dependent, for example one could trivially choose
\begin{equation*}
    N(E) = \begin{cases} \sum_{\mathbf{n}} P_{\mathbf{n},r}(E) & \text{ for } E<10 \text{kcal mol}^{-1} \\
   \sum_{\mathbf{n}} P_{\mathbf{n},p}(E) & \text{ for } E\geq10 \text{kcal mol}^{-1}. \end{cases}
\end{equation*}
These facts alone will be enough to understand Sec.~\ref{sec:General_Theory}, however, they can be illustrated  more concretely by a brief discussion of scattering theory, in particular the scattering matrix, $\mathbf{S}(E)$,\cite{TaylorScatteringBook} as is done in Appendix~\ref{app:Scattering}.

\section{General Theory} \renewcommand{\theequation}{3.\arabic{equation}}
\setcounter{equation}{0} \label{sec:General_Theory}
Following the discussion of the previous section we start by writing the cumulative reaction probability as
\begin{equation}
    N(E) = \sum_{\mathbf{n}} P_{\mathbf{n}}(E)
\end{equation}
where the probabilities $P_{\mathbf{n}}(E)$ are defined relative to some (as yet unspecified) energy dependent  basis. %
The central idea that underpins the unified semiclassical framework is that, 
under suitable (transition state theory like) assumptions [discussed further in Sec.~\ref{sec:One_dimension}], there exists a basis such that the probability is given \emph{exactly} by
\begin{equation}
    P_{\mathbf{n}}(E) = \frac{1}{1+e^{\tilde{W}_{\mathbf{n}}(E;\hbar)/\hbar}} \label{eq:Key_ansatz_1}
\end{equation}
where $\tilde{W}_{\mathbf{n}}(E;\hbar)$ is an effective ($\hbar$ dependent) action that behaves asymptotically as
\begin{equation}
    \tilde{W}_{\mathbf{n}}(E;\hbar) \sim W_{0}(E) + \sum_{n=1}^\infty W_{\mathbf{n},n}(E)\hbar^n \label{eq:Key_ansatz_2}
\end{equation}
as $\hbar\to0$. Here, $W_0(E)$ is again just the classical (Euclidean) reduced action along the instanton path with energy $E$ defined as
\begin{equation}
    W_0(E) = 2\int_{x_-}^{x_+} \sqrt{2\left(V(x)-E\right)} \, \mathrm{d}x \equiv  \oint \bm{p}\cdot\mathrm{d}\bm{q} \label{eq:classical_action_definition}
\end{equation}
where now, $x$ parameterises the mass-weighted position along the path between the turning points, $V(x_{\pm})=E$. Alternatively, this can be viewed as a line integral of the classical ``imaginary-time'' momentum, $\bm{p}$, along the full instanton path. 
 Physically, Eqs.~(\ref{eq:Key_ansatz_1}) and (\ref{eq:Key_ansatz_2}) imply that for each total energy, $E$, there exists a choice of basis such that as $\hbar\to0$ all the probabilities,  $P_{\mathbf{n}}(E)$,  are either zero or one (as one would expect from a ``classical'' limit). %
 Importantly, the form of Eq.~(\ref{eq:Key_ansatz_1}) is exactly equivalent to the  SCTST ansatz.\cite{Miller1977SCTST,Seideman1991SCTST_Siegert_eigvals} The key difference, therefore, between the present theory and SCTST is in how the action is defined.

To define $\tilde{W}_{\mathbf{n}}(E;\hbar)$ we consider the Green's function, $\hat{G}(E)=(E+i\epsilon^{+}-\hat{H})^{-1}$.\footnote{Here, $\epsilon^{+}$ is a positive infinitesimal.}  
Here, we use the fact that Gutzwiller's famous semiclassical expression for the trace of the Green's function\cite{GutzwillerBook,HellerBook} can be upgraded to an exact expression, such that the trace can be written exactly in terms of a sum over
all
classical ``prime periodic orbits'' (p.p.o.'s) as\cite{Ture2025ExactWKB,ChaosBookAppendixOnGreensFunction}
\begin{equation}
    \tr[\hat{G}(E)] =  \frac{1}{\hbar}\sum_{\sigma\in {\rm p.p.o.}} \sum_{\mathbf{n}} \sum^{\infty}_{r=1} (-1)^{r} \tilde{\tau}_{\sigma,\mathbf{n}}(E;\hbar) e^{-r \tilde{W}_{\sigma,\mathbf{n}}(E;\hbar)/\hbar} \label{eq:Greens_Fn}
\end{equation}
where $\sigma$ labels the periodic orbit.
The key difference from Gutzwiller's expression is that the action, $W$, and period, $\tau$, of each p.p.o.~are replaced by ``quantum'' counterparts.
That is to say, $\tilde{W}_{\sigma,\mathbf{n}}(E;\hbar)$ is a generalised ``quantum'' Euclidean action for the orbit, which also depends on  a set of quantum numbers, $\mathbf{n}$, of a new Schr\"odinger equation defined by the orbit, and $\tilde{\tau}_{\sigma,\mathbf{n}}(E;\hbar)= -\tilde{W}'_{\sigma,\mathbf{n}}(E;\hbar)$ is the generalised ``quantum'' period of the orbit. For the present purpose, the prime periodic orbit of interest is the instanton. Note that, throughout, we shall also refer to the continuation of the orbit at energies above the barrier as the instanton.

 Now both Eq.~(\ref{eq:Key_ansatz_1}) and Eq.~(\ref{eq:Greens_Fn}) involve quantum actions, it is, therefore, natural to suggest that (up to hyperasymptotic corrections) the exact action, $\tilde{W}_{\mathbf{n}}(E;\hbar)$, appearing in Eq.~(\ref{eq:Key_ansatz_1}) is equivalent to the generalised action of the instanton orbit, $\tilde{W}_{{\rm inst},\mathbf{n}}(E;\hbar)$, of Eq.~(\ref{eq:Greens_Fn}),
 \begin{equation}
     \tilde{W}_{\mathbf{n}}(E;\hbar)\sim\tilde{W}_{{\rm inst},\mathbf{n}}(E;\hbar) +\mathcal{O}(e^{-A/\hbar}).
 \end{equation}
 This conjecture provides an alternative practical route to the calculation of $\tilde{W}_{\mathbf{n}}(E;\hbar)$, that avoids any explicit reference to the concept of ``good'' action-angle variables of Miller's original SCTST\@.\cite{Miller1977SCTST}
 For example, 
 following Ref.~\citenum{ChaosBookAppendixOnGreensFunction}, one can derive explicit expressions for the $W_{\mathbf{n},n}(E)$ in terms of properties of the instanton orbit.
 Alternatively, one can obtain simple ring-polymer instanton expressions by matching terms with a standard path-integral calculation of the instanton contribution to the Green's function.
 
While it is obvious that there must be some connection between the instanton contribution to the Green's function and the rate problem, it is perhaps not obvious that they should be related in specifically this manner. 
We will leave a careful first-principles proof for future work, and in the following we will simply consider a series of examples that give credence to the assertion.

The first of these is that, consulting  Ref.~\citenum{ChaosBookAppendixOnGreensFunction}, one finds the first-order correction to the action is given by
\begin{equation}
    W_{\mathbf{n},1}(E)= \sum_{k=1}^{F-1} u_{k}(E)\left (n_k+\tfrac{1}{2}\right)
\end{equation}
where $\{u_k(E)\}$ are the stability parameters for the orbit that appear in Miller's original theory. Note that, for separable systems  $u_k(E)\equiv\tau(E)\omega_k$ where $\omega_k$ are the frequencies of the orthogonal (transverse) modes and $\tau(E)$ is the classical period of the instanton orbit. From this, as we shall show in detail in Sec.~\ref{sec:instanton_connection}, it follows that at leading order one recovers the standard instanton expression for $k(\tau)$.\cite{Miller1975semiclassical} This highlights that one needs to go to at least $W_{\mathbf{n},2}(E)$ in order to obtain corrections to instanton theory. We leave the explicit calculation of $W_{\mathbf{n},2}(E)$ in multiple dimensions to later papers, for now, however, we consider some simple results in one dimension.

\section{One dimension and exact WKB}\label{sec:One_dimension}\renewcommand{\theequation}{4.\arabic{equation}}
\setcounter{equation}{0}
In the absence of transverse degrees of freedom the sum over $\mathbf{n}$ disappears and we have
\begin{equation}
    N(E)=P(E)=\frac{1}{1+e^{\tilde{W}(E;\hbar)/\hbar}}.
\end{equation}
This, of course, is just a simple generalisation of the uniform WKB transmission probability [Eq.~(\ref{eq:uniform_WKB_PE})], in which the classical instanton action,  $W_0(E)$, has been replaced by $\tilde{W}(E;\hbar)$.
Following Ref.~\citenum{Upadhyayula2024hbar2corrections}, this can be used to define the exact action, $\tilde{W}(E;\hbar)$, in terms of the transmission probability
\begin{equation}
    \tilde{W}(E;\hbar) = \hbar \log(P(E;\hbar)^{-1}-1). \label{eq:1D_exact_W_in_terms_of_P_formula}
\end{equation}
This is formally useful in the analysis of $\tilde{W}(E;\hbar)$ in simple systems, however, as pointed out in Ref.~\citenum{Upadhyayula2024hbar2corrections} it is clearly limited by the need to know $P(E)$. 

Before we return to our definition of $\tilde{W}(E;\hbar)$ in terms of Eq.~(\ref{eq:Greens_Fn}), we note that we can use Eq.~(\ref{eq:1D_exact_W_in_terms_of_P_formula}) to gain insight into when $\tilde{W}(E;\hbar)$ [and hence $\tilde{W}_{\mathbf{n}}(E;\hbar)$] satisfies Eq.~(\ref{eq:Key_ansatz_2}). Clearly, in cases where the WKB transmission probability is correct asymptotically, that is if 
\begin{equation}
    \lim_{\hbar\to0} P(E;\hbar) \left[1+e^{W_0(E)/\hbar}\right] = 1,
\end{equation}
then the leading order term in our series for $\tilde{W}(E;\hbar)$ is necessarily the constant, $W_0(E)$. From this one can see that the validity of the asymptotic expansion  rests on the same assumptions as Kemble's uniform transmission probability. Hence, we require that the potential is analytic in a simply connected region containing the turning points. Further, following the analysis of Ref.~\citenum{Lawrence2024crossover}, we require that real-time trajectories starting at the two turning points of the instanton trajectory reach the reactants and products respectively, i.e.~there is  no complex forming or resonance phenomena. 
Finally, we assume that there is a single dominant instanton at each energy, and that the paths at different energies are continuously related.
Practically this final assumption can be relaxed provided the paths can be treated separately. However, as with all instanton methods, the theory will break down in liquids where a there are a large number of instantons that interact strongly. In such systems, methods such as ring-polymer molecular dynamics that involve direct sampling of the path-integral are to be preferred.\cite{RPMDrate,RPMDrefinedRate,Habershon2013RPMDreview,Lawrence2020rates}
Ultimately, from a qualitative perspective, we are making a generalised transition state theory approximation.
Following Ref.~\citenum{Ture2025ExactWKB}, we note that in one dimension the ``quantum'' reduced action appearing in Eq.~(\ref{eq:Greens_Fn}) is precisely the reduced action for the prime periodic orbit from ``exact WKB theory.'' 
Hence, under the conditions outlined in the previous paragraph we have that (up to hyper-asymptotic corrections) $\tilde{W}(E;\hbar)$ is the exact WKB reduced (Euclidean) action of the instanton. Exact WKB is a very rich and complex subject that is the result of applying the ideas of resurgent asymptotics\cite{Dingle1973AsymptoticsBook,Ecalle1981Resurgence,Dunne2014ResurgentTransseries} to semiclassical analysis of the Schr\"odinger equation. In what follows we give a surface level summary, introducing the ideas relevant to the present discussion, for  more details see e.g.~Ref.~\citenum{Ture2025ExactWKB} and references therein.\footnote{It is of note that similar ideas have been independently explored in the chemical physics literature: Y. Goldfarb, I. Degani and D. J. Tannor, ``Bohmian mechanics with complex action: A new trajectory based formulation of quantum mechanics,'' J. Chem. Phys. {\bf 125}, 231103 (2006); J. Schiff, Y. Goldfarb and D. J. Tannor, ``Path integral derivations of complex trajectory methods,'' Phys. Rev. A {\bf 83}, 012104 (2011); N. Zamstein and D. J. Tannor, ``Overcoming the root search problem in complex quantum trajectory calculations,'' J. Chem. Phys. {\bf 140}, 041105 (2014).}

The starting point of exact WKB is the WKB ansatz for the wavefunction
\begin{equation}
    \psi(q,E) = e^{-\tilde{W}(q,E;\hbar)/\hbar}
\end{equation}
which defines the position dependent action $\tilde{W}(q,E;\hbar)$.
Inserting this definition into the time-independent Schr\"odinger equation gives the standard non-linear Riccati equation for the log-derivative of the wavefunction\cite{Johnson1973logderivative,Mano1986logderivative}
\begin{equation}
    \tilde{p}^2(q;\hbar)=2m[V(q)-E]+\hbar \tilde{p}'(q;\hbar) \label{eq:Riccati}
\end{equation}
with the log-derivative defined by
\begin{equation}
    \tilde{p}(q;\hbar)= \tilde{W}'(q;\hbar).
\end{equation}
As can be seen from Eq.~(\ref{eq:Riccati}),  $\tilde{p}$ can naturally be interpreted as the ``quantum'' generalisation of the classical (imaginary-time) momentum.

The exact action for the instanton orbit then becomes
\begin{equation}
    \tilde{W}(E;\hbar)=\oint \tilde{p}(q,E;\hbar) \mathrm{d}q
\end{equation}
where the integral is performed around a closed loop in the complex plane that encloses the two turning points. [Note that by Cauchy's integral theorem the integral can be performed along any loop in the complex plane that encloses the branch points of $\tilde{p}(q,E;\hbar)$ corresponding to the two turning points.]
Expanding $\tilde{p}$ in Eq.~(\ref{eq:Riccati}) in powers of $\hbar$  and equating coefficients one can show that [in one dimension], after integrating around the closed loop,  only even terms survive, giving
\begin{equation}
    \tilde{W}(E;\hbar)\sim W_0(E) + \sum_{n=1}^\infty \hbar^{2n}W_{2n}(E) \label{eq:1D_W_expansion}
\end{equation}
where again $W_0(E)$ is the classical action. Following this procedure one can obtain explicit expressions for the $W_{2n}(E)$. For example, one finds the $\hbar^2$ correction is given by 
\begin{equation}
    W_2(E)=\frac{1}{2^5}  \oint \frac{(2mV'(q))^2}{\left(\sqrt{2m(V(q)-E)}\right)^5}\mathrm{d}q. \label{eq:Exact_W2_1D}
\end{equation}
 Note that the branch of the square root  must be chosen to be consistent with the integral for $W_0(E)$. A similar expression for $W_4(E)$ is given in Appendix~\ref{app:W_4}.

\subsection{Eckart barrier}
We can now test the conjecture of Sec.~\ref{sec:General_Theory} by comparing Eq.~(\ref{eq:1D_W_expansion}) with the expansion of Eq.~(\ref{eq:1D_exact_W_in_terms_of_P_formula}) for systems for which the exact transmission probability is known analytically. One such system is the asymmetric Eckart barrier, for which the potential has the form
\begin{equation}
    V(q) = \frac{\left(\sqrt{V_1}+\sqrt{V_2}\right)^2}{4\cosh^2(q/L)} -\frac{V_2-V_1}{1+\exp(-2q/L)}.
\end{equation}
As shown in Appendix~\ref{app:Eckart_Action}, inserting the exact expression for $P(E)$ into Eq.~(\ref{eq:1D_exact_W_in_terms_of_P_formula}) followed by a simple algebraic manipulation gives
\begin{equation}
    W_{2n}(E)= -2\pi\sqrt{\frac{4V_1V_2}{\omega^2}}  \frac{\omega^{2n}}{2^{6n}V_1^nV_2^n} \frac{(2n)!}{(n!)^2(2n-1)} \label{eq:Eckart_W2n}
\end{equation}
where $\omega$ is the (real number) unstable frequency at the barrier top.
It is then straightforward to confirm numerically that for $n=1$ and $n=2$ this is equivalent to the exact WKB expressions for $W_2(E)$ [Eq.~(\ref{eq:Exact_W2_1D})] and $W_4(E)$ [Eq.~(\ref{eq:W_4_1D})]. 
While still not a proof, this is a strong second piece of evidence in support of the conjecture.

The analytic form of Eq.~(\ref{eq:Eckart_W2n}) shows that the Eckart barrier is rather special. For example, one finds that 
the power series has a convergent representation which can be written in terms of $W_2(E)$ as\footnote{This is of course, therefore, an obvious choice of resummation scheme in more complex systems.}
\begin{equation}
    \sum_{n=1}^\infty \hbar^{2n}W_{2n}(E) = \frac{\pi^2}{2 W_2(E)} \left(1-\sqrt{1-4\hbar^2 W_2(E)^2/\pi^2}\right).
\end{equation}
Furthermore, we see that the Eckart barrier is a special case for which $W_{2n}(E)$ are independent of the energy, $E$. One might, therefore, be concerned that the agreement between our conjecture and the exact form is somehow a special feature of the Eckart barrier.

\subsection{Gaussian barrier}
\begin{figure}[t]
    \centering
    \includegraphics{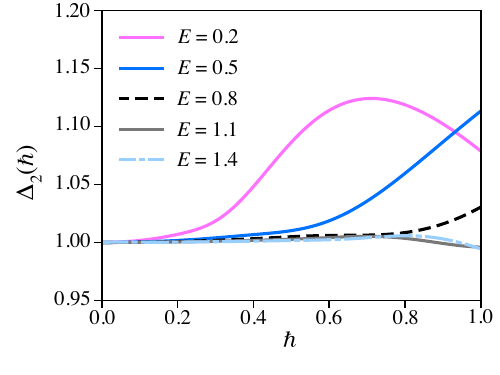}
    \caption{Graphical confirmation that $\lim_{\hbar\to0}[\tilde{W}(E;\hbar)-W_0(E)]/\hbar^2=W_2(E)$ for the Gaussian barrier defined by Eq.~(\ref{eq:Gaussian_potential}). The plotted ratios were calculated on a grid between $\hbar=0.1$ and $
    \hbar=1.0$, and cubic splines are used to guide the eye. }
    \label{fig:One_dimension_gaussian}
\end{figure}

To provide further test of the conjecture we, therefore, consider the Gaussian barrier, with potential
\begin{equation}
    V(q)=e^{-q^2/2} \label{eq:Gaussian_potential}
\end{equation}
and $m=1$.
While there is not a simple analytical solution for $P(E)$ for this system, we can nevertheless obtain $P(E)$ to high accuracy using the log-derivative method for comparison.\cite{Johnson1973logderivative,Mano1986logderivative} Fig.~\ref{fig:One_dimension_gaussian} shows 
\begin{equation}
    \Delta_2(\hbar) =\frac{\tilde{W}(E;\hbar)-W_0(E)}{\hbar^2 W_2(E)} \label{eq:Delta2}
\end{equation}
as a function of $\hbar$ for a series of different values of $E$ above and below the barrier height, with $\tilde{W}(E;\hbar)$ calculated using Eq.~(\ref{eq:1D_exact_W_in_terms_of_P_formula}) and $W_2(E)$ calculated using Eq.~(\ref{eq:Exact_W2_1D}). We see that $\Delta_2(\hbar)\to1$ as $\hbar\to0$, which demonstrates that Eq.~(\ref{eq:Exact_W2_1D}) is the correct asymptotic coefficient at $\hbar^2$ for this system, further supporting the conjecture.

\subsection{Separable systems}
Finally we note that if the conjecture is valid for one-dimensional systems then it is trivial to prove that it also holds for separable multidimensional systems, in which the reaction coordinate is uncoupled from the other degrees of freedom. This can be seen by noting that for such a system the cumulative reaction probability can be written as a convolution of the 1D transmission probability of the reaction coordinate, $P(E;\hbar)$, with the density of states in the orthogonal (transverse) degrees of freedom, $\rho_\perp(E;\hbar)$, as
\begin{equation}
    N(E) = \int_{-\infty}^\infty  P(E_{\rm rc};\hbar)\rho_\perp(E-E_{\rm rc};\hbar)\, \mathrm{d}E_{\rm rc}. \label{eq:sep_convolution}
\end{equation}
Then, we can use the fact that for separable systems the instanton contribution to the Green's function in Eq.~(\ref{eq:Greens_Fn}) can trivially be written in terms of the 1D action $\tilde{W}(E;\hbar)$ as 
\begin{equation}
\begin{aligned}
    \left(\tr[\hat{G}(E)]\right)_{\rm inst} &=  \frac{-1}{\hbar} \sum_{\mathbf{n}}   \tilde{\tau}_{\mathbf{n}}(E;\hbar) e^{- \tilde{W}_{\mathbf{n}}(E;\hbar)/\hbar} \\
    &= \frac{-1}{\hbar} \sum_{\mathbf{n}}   \tilde{\tau}(E-\tilde{E}^\perp_{\mathbf{n}}(\hbar);\hbar) e^{- \tilde{W}(E-\tilde{E}^\perp_{\mathbf{n}}(\hbar);\hbar)/\hbar}
    \end{aligned}
\end{equation}
where $\tilde{E}^\perp_{\mathbf{n}}(\hbar)$ are the energy eigenvalues for the degrees of freedom orthogonal to the reaction. Hence, applying the conjecture one recovers the correct form [Eq.~(\ref{eq:sep_convolution})] with a couple of simple manipulations
\begin{equation}
\begin{aligned}
    N(E) &= \sum_{\mathbf{n}} \frac{1}{1+e^{\tilde{W}(E-\tilde{E}^\perp_{\mathbf{n}}(\hbar);\hbar)/\hbar}}\\
    &=\int_{-\infty}^\infty  \frac{1}{1+e^{\tilde{W}(E_{\rm rc};\hbar)/\hbar}} \sum_{\mathbf{n}}\delta(E-\tilde{E}^\perp_{\mathbf{n}}(\hbar)-E_{\rm rc}) \, \mathrm{d}E_{\rm rc}
    \\
    &=\int_{-\infty}^\infty P(E_{\rm rc};\hbar)\rho_\perp(E-E_{\rm rc};\hbar)  \, \mathrm{d}E_{\rm rc}.
\end{aligned}
\end{equation}

\section{Connection to Vibrational Perturbation Theory and SCTST}\label{sec:SCTST} \renewcommand{\theequation}{5.\arabic{equation}}
\setcounter{equation}{0}

To make connection with earlier work we now turn to consider vibrational perturbation theory (VPT) and SCTST\@.\cite{Green1990VPT2,Miller1990SCTST} In the following we will recast Miller's VPT-SCTST approach in the present notation, giving a new perspective on the approximations made in the standard VPT2-SCTST method,\cite{Greene2016SCTST,Nguyen2010SCTST,Burd2018SCTSTTorsion,Mandelli2022SCTSTOrganicMolecules,Aieta2019SCTST,Conte2024PerspectiveSCTST_and_Spectra} and allowing us to define key quantities that will be useful when we discuss the connection to instanton theory in Sec.~\ref{sec:instanton_connection}.
Before discussing SCTST, however, we begin with a brief recap of the idea behind VPT, emphasising the connection of $\hbar$ to the perturbation parameter.

\subsection{Summary of VPT}
Working in mass weighted normal mode coordinates we have that the Hamiltonian for an $F$ dimensional system,
\begin{equation}
    \hat{H} = \sum_{j=1}^{F} \frac{\hat{P}_j^2}{2} + V(\hat{\bm{Q}}),
\end{equation}
can be expanded about a minimum as
\begin{equation}
    \hat{H} = V_{\rm min}+ \sum_{j=1}^{F} \frac{\hat{P}_j^2}{2}  +\sum_{j=1}^F \frac{\omega_j^2 \hat{Q}_j^2}{2} + \sum_{n=3}^\infty\sum_{j_1,\dots,j_n} \frac{f_{j_1,\dots,j_n}}{n!}\prod_{k=1}^n \hat{Q}_{j_k}
\end{equation}
where the force constants are defined by $f_{j_1,j_2,j_3}=\frac{\partial^3V}{\partial Q_{j_1}\partial Q_{j_2}\partial Q_{j_3}}$.
To arrive at the necessary form for vibrational perturbation theory one then simply rewrites this in terms of harmonic oscillator raising and lowering operators
and divides out by $\hbar$
to give
\begin{equation}
    \frac{\hat{H}}{\hbar} = \frac{V_{\rm min}}{\hbar}+ \sum_{j=1}^{F} \omega_j(\hat{a}_j^\dagger\hat{a}_j+\tfrac{1}{2}) + \sum_{n=1}^\infty \hbar^{\frac{n}{2}} \lambda^n \hat{U}_n,
\end{equation}
where we have introduced the perturbation theory parameter $\lambda$, and defined the operators as
\begin{equation}
    \hat{U}_{n-2}=\sum_{j_1,\dots,j_n}  \frac{f_{j_1,\dots,j_n}}{2^{n/2}n!}\prod_{k=1}^n \frac{\hat{a}_{j_k}\!+\!\hat{a}^\dagger_{j_k}}{\sqrt{\omega_{j_k}}}.
\end{equation}
We thus see that $n^{\rm th}$ order perturbation theory can be directly associated with a factor of $\hbar^{n/2}$, and we can simply use $\hbar$ in place of the book-keeping parameter $\lambda$. The vibrational perturbation theory expansion of the energy is therefore equivalent to an asymptotic expansion of the energy in $\hbar$. As expectation values of $\hat{U}_1$ with eigenstates of the reference Hamiltonian are zero there is no correction at first order. The energy is thus given perturbatively up to order $\hbar^2$ as
\begin{equation}
    E \!\sim\! V_{\rm min} + \sum_{j=1}^{F} \hbar\omega_j (n_j+\tfrac{1}{2}) +\hbar^2\gamma_0+ \sum_{j\leq j'} \hbar^2 \chi_{jj'}(n_j+\tfrac{1}{2})(n_{j'}+\tfrac{1}{2}) + \dots
\end{equation}
where $\gamma_0$ is a zero-point energy correction. For completeness we give the expressions for the VPT2 constants, $\gamma_0$ and $\chi_{jj'}$, in the current notation in Appendix \ref{app:VPT2_constants}.

\subsection{VPT-SCTST}
The basis of the VPT-SCTST method is the idea that for reaction rates one can simply apply VPT to the saddle point of the potential, analytically continuing the frequency of the unstable mode, which we take to be labelled $F$, and identifying its action with the barrier penetration action.\cite{Miller1990SCTST} In our current notation this is equivalent to saying 
\begin{equation}
    2\pi i\hbar\left(n_{F}+\tfrac{1}{2}\right)\Rightarrow \tilde{W}_{\mathbf{n}}(E;\hbar) \label{eq:SCTST_prescription}
\end{equation}
i.e.~our ``exact'' action is (at least asymptotically) equivalent to Miller's ``good'' action variable for the barrier transmission.\cite{Miller1977SCTST}$^,$\footnote{
Despite the term ``good'', as noted previously,\cite{Seideman1991SCTST_Siegert_eigvals} the use of perturbation theory means that the action does not actually have to be ``good'', i.e.~the system does not have to be integrable. Here, we note further that the generalisation of Bohr-Sommerfeld quantisation to chaotic systems\cite{ChaosBookAppendixOnGreensFunction} means semiclassical framework can be retained even in strongly coupled (non-perturbative) systems.} 

We now turn to what information one obtains about $\tilde{W}_{\mathbf{n}}(E;\hbar)$ from a given order of VPT-SCTST, and in particular from VPT2. Applying the SCTST prescription [Eq.~(\ref{eq:SCTST_prescription})] to the full perturbation series results in an expansion of the energy in powers of the action of the form
\begin{equation}
    E =  V^\ddagger + \sum_{\nu=0}^\infty a_{\mathbf{n},\nu}(\hbar) \tilde{W}^\nu_{\mathbf{n}}(E;\hbar).
\end{equation}
Defining $\tilde{V}_{\mathbf{n}}(\hbar)=V^\ddagger+a_{\mathbf{n},0}(\hbar)$ we can subtract $\tilde{V}_{\mathbf{n}}(\hbar)$ from both sides to obtain an invertible expansion.  Performing the inversion term by term (or using Lagrange inversion) then  gives an expansion for the action in powers of $E - \tilde{V}_{\mathbf{n}}(\hbar)$ of the form 
\begin{equation}
    \tilde{W}_{\mathbf{n}}(E;\hbar) = \frac{1}{a_{\mathbf{n},1}(\hbar)}(E - \tilde{V}_{\mathbf{n}}(\hbar)) - \frac{a_{\mathbf{n},2}(\hbar)}{a_{\mathbf{n},1}^3(\hbar)}(E -\tilde{V}_{\mathbf{n}}(\hbar))^2 + \dots \label{eq:inverted_VPT_SCTST}
\end{equation}
which is equivalent to (the first two terms in) the expansion
\begin{equation}
    \tilde{W}_{\mathbf{n}}(E;\hbar) = \sum_{m=0}^\infty \frac{\tilde{W}_{\mathbf{n}}^{(m)}(\tilde{V}_{\mathbf{n}}(\hbar);\hbar)}{m!}(E - \tilde{V}_{\mathbf{n}}(\hbar))^m. \label{eq:Action_expansion_around_V_tilde}
\end{equation}
Note that, by analogy with the equation $W_0(V^\ddagger)=0$, we can consider $\tilde{V}_{\mathbf{n}}(\hbar)$ as a generalised barrier height, as it is the energy at which $\tilde{W}_{\mathbf{n}}(\tilde{V}_{\mathbf{n}}(\hbar);\hbar)=0$.

Crucially, at a given order in VPT we do not obtain any of the $a_{\mathbf{n},n}(\hbar)$ to all orders in $\hbar$.
To see why this is so, note that, in $n^{\rm th}$ order perturbation theory, application of Eq.~(\ref{eq:SCTST_prescription}) results in terms up to $\tilde{W}_{\mathbf{n}}^{n/2+1}$. However, each application of Eq.~(\ref{eq:SCTST_prescription}) also removes a power of $\hbar$. Hence, at $n^{\rm th}$ order in perturbation theory the $\nu^{\rm th}$ order term in the action is associated with terms up to $\hbar^{n/2+1-\nu}$.
Therefore, defining 
\begin{equation}
    a_{\mathbf{n},\nu}(\hbar) = \sum_{m=0}^\infty a_{\mathbf{n},\nu,m}\hbar^m \label{eq:a_n_nu_expansion}
\end{equation}
we can see that from VPT2 we obtain $a_{\mathbf{n},\nu,m}$ for $\nu=0$ to $2$ and $m=0$ to $(2-\nu)$, i.e.~$a_{\mathbf{n},0,0}=0$, $a_{\mathbf{n},0,1}$, $a_{\mathbf{n},0,2}$, $a_{\mathbf{n},1,0}$, $a_{\mathbf{n},1,1}$ and $a_{\mathbf{n},2,0}$. 
In the original SCTST approach one assumes all other $a_{\mathbf{n},\nu,m}=0$.\cite{Miller1990SCTST} 
This is equivalent to assuming that the action can be resummed using the classical action of an inverted Morse oscillator.\cite{Wagner2013SCTST}

To determine what properties of $\tilde{W}_{\mathbf{n}}(E;\hbar)$ are obtained at the level of VPT2, one expands the coefficients of $(E-\tilde{V}_{\mathbf{n}})$ in terms of $\hbar$ in both Eqs.~(\ref{eq:inverted_VPT_SCTST}) and (\ref{eq:Action_expansion_around_V_tilde})  and equates powers. This is done explicitly in Appendix~\ref{app:W_from_VPT2} from which one finds
\begin{subequations}\label{eq:VPT2_W_identities}
    \begin{equation}
        W_0(V^\ddagger) = 0
    \end{equation}
    \begin{equation}
        W'_0(V^\ddagger) = -\frac{2\pi}{\omega} = -\tau_{\rm c}
    \end{equation}
    \begin{equation}
        W''_0(V^\ddagger) = -\frac{4\pi \chi_{F\!F}}{\omega^3}
    \end{equation}
    \begin{equation}
        W_{\mathbf{n},1}(V^\ddagger) = \tau_{\rm c}V_{\mathbf{n},1}
    \end{equation}
    \begin{equation}
        W'_{\mathbf{n},1}(V^\ddagger) = -\frac{2\pi \sum_{k=1}^{F-1} \chi_{kF} \left(n_k+\tfrac{1}{2}\right)}{\omega^2} + \frac{4\pi \chi_{F\!F}}{\omega^3}V_{\mathbf{n},1} \label{eq:W_n1_first_deriv}
    \end{equation}
        \begin{equation}
           W_{\mathbf{n},2}(V^\ddagger)= \frac{2\pi \sum_{k=1}^{F-1} \chi_{kF} \left(n_k+\tfrac{1}{2}\right)}{\omega^2} V_{\mathbf{n},1}+\frac{2\pi V_{\mathbf{n},2}}{\omega} -\frac{2\pi \chi_{F\!F}}{\omega^3}V_{\mathbf{n},1}^2 .
\end{equation}
\end{subequations}
where we have defined the expansion coefficients of $\tilde{V}_{\mathbf{n}}(\hbar)$ as
\begin{subequations}
\begin{equation}
    V_{\mathbf{n},1} = \sum_{k=1}^{F-1}  \omega_k\left(n_k+\tfrac{1}{2}\right)
\end{equation}
\begin{equation}
    V_{\mathbf{n},2} = \gamma_0 + \sum_{k'\leq k=1}^{F-1} \chi_{kk'} \left(n_k+\tfrac{1}{2}\right)\left(n_{k'}+\tfrac{1}{2}\right)
\end{equation}
\end{subequations}
and, as is standard, the VPT2 constants have been redefined  to avoid the appearance of complex numbers, see Appendix~\ref{app:SCTST/VPT2_constants} for full definitions. Note also $\omega=|\omega_F|$ is the magnitude of the imaginary frequency. 
Hence, we see that VPT2 determines all of $W_{0}^{}(V^\ddagger)$, $W_{0}^{\prime}(V^\ddagger)$, $W_{0}^{\prime\prime}(V^\ddagger)$, $W_{\mathbf{n},1}^{}(V^\ddagger)$, $W_{\mathbf{n},1}^{\prime}(V^\ddagger)$, and $W_{\mathbf{n},2}^{}(V^\ddagger)$. Crucially, VPT2 contains this and \emph{only} this information. Any additional terms given by the standard VPT2-SCTST method are a result of the particular choice of resummation used. While the resummation must obey certain properties, such as being size consistent, there is still a great degree of flexibility in the form used. Potential improvements to the standard scheme will be explored in following papers.

Before continuing we pause to highlight that Eq.~(\ref{eq:W_n1_first_deriv}) implies that the effective instanton frequencies defined by dividing the stability parameters by the period of the orbit, $\omega_k(E)=u_k(E)/\tau(E)$,\cite{Chapman1975rates} are related to the anharmonicity constants of VPT2-SCTST according to  
\begin{equation}
    -\frac{\chi_{kF}}{\omega} = \omega_k'(V^\ddagger).
\end{equation}
We will leave a detailed consideration of the  microcanonical instanton theory approach of Chapman, Garrett and Miller to a following work. However, we note that this behaviour fits naturally with their ideas.

\section{Connection to thermal instanton rate theory} \renewcommand{\theequation}{6.\arabic{equation}}
\setcounter{equation}{0}\label{sec:instanton_connection}
In this section we will consider how thermal instanton rate theory\cite{Miller1975semiclassical,Coleman1977ImF,Affleck1981ImF,InstReview} is derived within the present framework. 
 So far we have focussed on the cumulative reaction probability $N(E)$ and in particular on the asymptotic expansion of the reduced action $\tilde{W}_{\mathbf{n}}(E;\hbar)$ in powers of $\hbar$, for which $W_0(E)$ is the leading order term. Instanton rate theory is an asymptotic expansion of the rate constant $k(\tau)$ in $\hbar$,
 \begin{equation}
    k(\tau)Z_r(\tau)\sim (kZ_{r})_{\rm inst,0}(1+\alpha_{\mathrm{inst},1}\hbar+\alpha_{\mathrm{inst},2}\hbar^2+\dots)
\end{equation}
 where the standard theory\cite{Miller1975semiclassical,Coleman1977ImF,Affleck1981ImF,InstReview} is just the leading order term in this series, $(kZ_{r})_{\rm inst,0}$. While it is more usual in chemistry and physics to refer to the inverse temperature $\beta=1/(k_{\rm B}T)$ than the thermal time $\tau=\beta\hbar$, it is important to stress that, from the point of view of semiclassics, $\tau$ is the more natural quantity to work with. 
 Crucially this has a practical significance because when we take $\hbar\to0$ we will obtain \emph{different} results if we treat $\beta$ as fixed or $\tau$ as fixed. Choosing $\tau$ as fixed is the choice that is consistent with earlier work\cite{Miller1975semiclassical,Coleman1977ImF,Affleck1981ImF,InstReview}  and with associating the perturbation parameter with the $\hbar$ in the phase factor $e^{iS/\hbar}$ in the path-integral formulation of quantum mechanics.\cite{Kleinert}

To derive instanton theory we begin by noting that $E=\tilde{V}_{\mathbf{n}}$ marks the boundary between the two convergent series representations for the transmission probability
\begin{equation}
    P_{\mathbf{n}}(E) = \begin{cases} \sum_{n=1}^\infty (-1)^{n+1} e^{-n\tilde{W}_\mathbf{n}(E;\hbar)/\hbar} & \text{ when } E\leq\tilde{V}_{\mathbf{n}}(\hbar) \\
    \sum_{n=0}^\infty (-1)^{n} e^{n\tilde{W}_\mathbf{n}(E;\hbar)/\hbar} & \text{ when } E\geq\tilde{V}_{\mathbf{n}}(\hbar).
    \end{cases}
\end{equation}
We then insert this into the definition of the thermal rate given by Eq.~(\ref{eq:k_N_E_relation}) to obtain
\begin{equation}
    \begin{aligned}
        kZ_r &= \frac{1}{2\pi\hbar} \sum_{\mathbf{n}} \int_{-\infty}^{\tilde{V}_\mathbf{n}} e^{-\tau E/\hbar} \sum_{n=1}^\infty (-1)^{n+1} e^{-n\tilde{W}_\mathbf{n}(E;\hbar)/\hbar} \, \mathrm{d}E \\
        &+ \frac{1}{2\pi\hbar} \sum_{\mathbf{n}} \int_{\tilde{V}_\mathbf{n}}^{\infty} e^{-\tau E/\hbar} \sum_{n=0}^\infty (-1)^{n} e^{n\tilde{W}_\mathbf{n}(E;\hbar)/\hbar} \, \mathrm{d}E. \label{eq:full_rate_sums}
    \end{aligned}
\end{equation}
Making a simple variable change to remove the $\mathbf{n}$ and $\hbar$ dependence from the integration limits we can reorder the sums and integrals to give
\begin{equation}
    \begin{aligned}
        kZ_r &= \sum_{n=1}^\infty \frac{(-1)^{n+1} }{2\pi\hbar}\!  \int_{-\infty}^{V^\ddagger}\! \sum_{\mathbf{n}}  e^{-\tau E/\hbar-\tau\Delta\tilde{V}_{\mathbf{n}}/\hbar-n\tilde{W}_\mathbf{n}(E+\Delta\tilde{V}_{\mathbf{n}};\hbar)/\hbar} \, \mathrm{d}E \\
        &+ \sum_{n=0}^\infty \frac{(-1)^{n}}{2\pi\hbar} \!  \int_{V^\ddagger}^\infty \! \sum_{\mathbf{n}} e^{-\tau E/\hbar-\tau\Delta\tilde{V}_{\mathbf{n}}/\hbar+n\tilde{W}_\mathbf{n}(E+\Delta\tilde{V}_{\mathbf{n}};\hbar)/\hbar} \, \mathrm{d}E  \label{eq:transformed_integral_rep}
    \end{aligned}
\end{equation}
where $\Delta\tilde{V}_{\mathbf{n}}=\tilde{V}_{\mathbf{n}}-V^\ddagger$.
To obtain the asymptotic series in $\hbar$ for $kZ_r$ one then simply follows standard asymptotic analysis\cite{BenderBook}  expanding the integrands in a series in $\hbar$ and integrating. For $\tau>2\pi /\omega$ the first integral in the first sum dominates asymptotically. Upon expanding the integrand to zeroth order in $\hbar$ then gives
\begin{equation}
    kZ_r\sim\frac{1}{2\pi\hbar}   \int_{-\infty}^{V^\ddagger} e^{-\tau E/\hbar-W_0(E)/\hbar}  Z
    (E;\tau) \left[1+\mathcal{O}(\hbar) \right]  \mathrm{d}E \label{eq:instanton_E_integral}
\end{equation}
where
\begin{equation}
    Z(E;\tau) =  \sum_{\mathbf{n}} e^{-\tau{V}_{\mathbf{n},1}-\left[W_{\mathbf{n},1}(E)+W_0'(E)V_{\mathbf{n},1}\right]}. \label{eq:generalised_partition_function_1}
\end{equation}
It is notable that this integrand is subtly different from that in Miller's original derivation\cite{Miller1975semiclassical} due to the terms involving $V_{\mathbf{n},1}$ in $Z(E;\tau)$. However, once the integral is evaluated by steepest descent (as $\hbar\to0)$  we obtain, at leading order,
\begin{equation}
    kZ_r\sim (kZ_{r})_{\rm inst,0} =\frac{1}{2\pi\hbar} \sqrt{\frac{2\pi\hbar}{W''_0(E^\star)}}e^{-\tau E^\star/\hbar-W_0(E^\star)/\hbar} Z(E^\star;\tau)
\end{equation}
which is equivalent to Miller's original expression,\footnote{As always this can be re-expressed in precisely the form Miller gives by using the relation of the reduced action to the total action, $S_{\! \rm inst}(\tau)=W_0(E^\star)+\tau E^\star$, along with $W_0''(E^\star)=-\left(\frac{\mathrm{d}E^\star}{\mathrm{d}\tau}\right)^{-1} $}  where the stationary point is defined by $W'_0(E^\star)=-\tau$. This is because the $V_{\mathbf{n},1}$ dependence of $Z(E^\star;\tau)$ exactly cancels at $E^\star$ to give
\begin{equation}
\begin{aligned}
    Z(E^\star;\tau)&=\sum_{\mathbf{n}} e^{- W_{\mathbf{n},1}(E^\star)} \\&= \sum_{\mathbf{n}} e^{- \sum_k u_k(E^\star) \left(n_k+\tfrac{1}{2}\right)}\\ &=\prod_{k=1}^{F-1}\frac{1}{2\sinh(u_k(E^\star)/2)} = Z_{\rm inst}(E^\star).
    \end{aligned} \label{eq:Z_instE}
\end{equation}

Of course, one of the advantages of the present formalism is that it makes the derivation of corrections to instanton theory straightforward. For example, as detailed in Appendix~\ref{app:instanton_connection}, by expanding the integrand in Eq.~(\ref{eq:instanton_E_integral}) up to first order in $\hbar$ and  applying standard asymptotic analysis the first-order correction can be shown to be given by
\begin{equation}
\begin{aligned}
    \!\!\alpha_{\mathrm{inst},1} \!&=\! \frac{Z''_{\rm inst}(E^\star)}{2 Z_{\rm inst}(E^\star) W_0''(E^\star)} - \frac{W_0'''(E^\star)Z'_{\rm inst}(E^\star)}{2Z_{\rm inst}(E^\star)[W_0''(E^\star)]^2} \\ &-\frac{W_0''''(E^\star)}{8[W_0''(E^\star)]^{2}}+\frac{5[W_0'''(E^\star)]^2}{24[W_0''(E^\star)]^{3}} - \langle W_{\mathbf{n},2}(E^\star)\rangle.
    \end{aligned}
\end{equation}
where again we used cancellation [akin to Eq.~(\ref{eq:Z_instE})] to write in terms of $Z_{\rm inst}(E^\star)$ rather than $Z(E^\star;\tau)$. Here the expectation value is defined as
\begin{equation}
    \langle W_{\mathbf{n},2}(E^\star)\rangle = \frac{\sum_{\mathbf{n}}e^{-W_{\mathbf{n},1}(E^\star)} W_{\mathbf{n},2}(E^\star)}{Z_{\rm inst}(E^\star)}.
\end{equation}

These formulas give the first-order corrected instanton rate in terms of properties of the classical instanton trajectory, such as the stability parameters, $u_k(E)$. However, using the conjectured connection between the generalised reduced action $\tilde{W}_{\mathbf{n}}(E;\hbar)$ and the Green's function these can alternatively be written in terms of discretised path-integrals using the standard ring-polymer instanton framework.\cite{InstReview} The present theory therefore enables the derivation of dividing-surface-independent formulations of the RPI+PC method,\cite{Dusek2025RPI+PC} which will be the subject of future work.

\section{Semiclassical rates above the crossover temperature: The ``sphaleron'' rate}\label{sec:sphaleron}\renewcommand{\theequation}{7.\arabic{equation}}
\setcounter{equation}{0}
The standard instanton rate theory is only valid at temperatures below the ``crossover temperature'' for which $\tau>2\pi/\omega$. Above this temperature the asymptotic series takes a different form and the rate is dominated by a trajectory localised at the transition state, i.e.~the activated complex, or, in the language of high energy physics, the ``sphaleron''.\cite{Klinkhamer1984Sphaleron} For $\tau<2\pi/\omega$ all of the integrals in Eq.~(\ref{eq:transformed_integral_rep}) contribute asymptotically (as $\hbar\to0$). Expanding these integrals up to leading order and integrating asymptotically (see Appendix~\ref{app:sphaleron_derivation}) one obtains 
\begin{equation}
    kZ_r\sim  \frac{1}{2\pi} e^{-\tau {V}^\ddagger/\hbar} \frac{{\omega}}{2\sin({\omega}\tau/2)} Z_h^{\ddagger}(\tau) \equiv k_{\text{m-pb}}Z_r
\end{equation}
with
\begin{equation}
    Z_{h}^\ddagger(\tau) = \prod_{k=1}^{F-1}\frac{1}{2\sinh(
    \tau\omega_k/2)}
\end{equation}
i.e.~the well known multidimensional parabolic barrier, harmonic oscillator approximation to the rate. 

Of course we can again go beyond this approximation to include higher order terms in the asymptotic series
\begin{subequations}
    \begin{equation}
    kZ_r\sim k_{\text{m-pb}}Z_r(1+\alpha_{{\rm sph},1}\hbar +\alpha_{{\rm sph},2}\hbar^2+\dots).
\end{equation}
Again expanding the integrands in Eq.~(\ref{eq:transformed_integral_rep}) up to first order and integrating asymptotically we obtain
\begin{equation}
    \alpha_{{\rm sph},1}(\tau) = \eta(\tau) - W''_0(V^\ddagger)G(\tau,\omega)-\tau\langle V_{\mathbf{n},2}\rangle
\end{equation}
\end{subequations}
where now the expectation value is over the harmonic transition state partition function
\begin{equation}
    \langle A_{\mathbf{n}}\rangle = \frac{\sum_{\mathbf{n}}e^{-\tau{V}_{\mathbf{n},1}} A_{\mathbf{n}}}{Z_{h}^\ddagger(\tau)}
\end{equation}
and we have defined
\begin{equation}
    \eta(\tau) =\frac{\omega}{2\pi}  \left(1-\frac{\tau  \omega/2}{\tan(\tau  \omega /2)}\right) \left[\langle W_{\mathbf{n},1}'(V^\ddagger)\rangle+W_0''(V^\ddagger)\langle {V}_{\mathbf{n},1} \rangle\right]
\end{equation}
and 
\begin{equation}
    G(\tau,{\omega})=  \frac{{\omega}^2\left(\tau{\omega}[3+\cos({\omega}\tau)]-4\sin({\omega}\tau)\right)}{32\pi \sin^2({\omega}\tau/2)}. \label{eq:G_tau_omega}
\end{equation}

Inserting the VPT2 identities, Eq.~(\ref{eq:VPT2_W_identities}), this can be written (in the absence of rotations) in terms of the VPT2-SCTST constants (defined in Appendix~\ref{app:SCTST/VPT2_constants}) as 
\begin{equation}
    \alpha_{\rm sph,1}(\tau) = \eta(\tau) + \frac{4\pi\chi_{F\!F}}{\omega^3}G(\tau,\omega) -\tau[\gamma_0 + \zeta(\tau) + \xi(\tau)] \label{eq:1st_order_sphaleron}
\end{equation}
where
\begin{subequations}
    \begin{equation}
    \eta(\tau) = \left(\frac{\omega\tau/2}{\tan(\omega\tau/2)}-1\right) \sum_{k=1}^{F-1}\frac{\chi_{kF}}{\omega}\tfrac{1}{2}\coth(\omega_k\tau/2)
\end{equation}
\begin{equation}
    \zeta(\tau)=  \sum_{k'< k=1}^{F-1} \chi_{kk'} \tfrac{1}{2}\coth(\omega_k\tau/2)\tfrac{1}{2}\coth(\omega_{k'}\tau/2)
\end{equation}
\begin{equation}
    \xi(\tau) =  \sum_{k=1}^{F-1} \frac{\chi_{kk}}{8}\frac{3+\cosh(\omega_k\tau)}{\sinh^2(\omega_k\tau/2)}.
\end{equation}
\end{subequations}

We can make connection between this result and earlier work in one dimension by Pollak and Cao that considered fixed $\beta$ as $\hbar\to0$,\cite{Pollak2022hbarSquared_Correction,Pollak2024hbar4_expansion} rather than fixed $\tau$ as $\hbar\to0$. 
Making the substitution $\tau=\beta\hbar$ in Eq.~(\ref{eq:1st_order_sphaleron}) one finds that (upon expanding each term again in $\hbar$) one recovers the full $\hbar^2$ correction derived in Ref.~\citenum{Pollak2022hbarSquared_Correction}   (consistent with the analysis of the SCTST rate\cite{Hernandez1993SCTST} made in that work\cite{Pollak2022hbarSquared_Correction}). From the present analysis it is also clear why it is not sufficient to use the $\hbar^2$ (VPT4) fixed $\tau$ expansion, to recover the next ($\hbar^4$) term in the fixed $\beta$ expansion,\cite{Pollak2024hbar4_expansion} as terms that appear at $\hbar^3$ in fixed $\tau$ will also contribute to the $\hbar^4$ with fixed $\beta$. Finally we note that the reverse process does not work, i.e.~one cannot start with a truncated fixed $\beta$ expansion in $\hbar$ and substitute $\beta=\tau/\hbar$ to get the fixed $\tau$ expansion. This is because, while going from $\tau$ to $\beta\hbar$ increases the associated power of $\hbar$, going from $\beta$ to $\tau/\hbar$ lowers it, and hence one would need the fixed $\beta$ expansion to all orders to recover the fixed $\tau$ expansion.

\section{Multidimensional anharmonic TST and anharmonic Wigner tunneling correction }\label{sec:AnhTST_Wigner}\renewcommand{\theequation}{8.\arabic{equation}}
\setcounter{equation}{0}
Despite its theoretical importance, the parabolic barrier approximation is not often used practically due to its unphysical divergence at the crossover temperature. It is, therefore, more common to use the harmonic Wigner tunneling correction\cite{Wigner1932parabolic}  
\begin{equation}
    \frac{\omega}{4\pi\sin(\omega\tau/2)} \Rightarrow \frac{1}{2\pi\tau}\kappa_{\rm hW} = \frac{1}{2\pi\tau}\left(1+\frac{\tau^2\omega^2}{24} \right),
\end{equation}
which, although it still becomes less accurate at low temperature, does not diverge unphysically. For the same reason, although the first-order (sphaleron) rate  derived in the previous section [Eq.~(\ref{eq:1st_order_sphaleron})] is theoretically useful,  on its own it is not of great practical utility in most of chemistry due to the divergence at the crossover temperature. While it is possible to overcome this divergence by combining it with the instanton rate (extending the work of Ref.~\citenum{Lawrence2024crossover} to first order), a detailed discussion of this approach and numerical examples is left for future work.
For now we note that in many chemical applications an accurate description of deep tunneling is not required.
Here, therefore, we focus instead on a simple multidimensional and anharmonic version of TST and the Wigner tunneling correction.

We start by noting that in the high temperature limit the most important effect is the change to the barrier height. Hence, a sensible leading order approximation to the rate is given by replacing the individual transmission probabilities with the step function approximation to the logistic function, 
\begin{equation}
    N(E) \approx \sum_{\mathbf{n}} \theta(E-\tilde{V}_{\mathbf{n}})
\end{equation}
i.e.~ignoring ``tunneling.'' The thermal rate is then simply
\begin{equation}
    k(\tau)Z_r(\tau) \approx \frac{1}{2\pi\tau}\sum_{\mathbf{n}}e^{-\tau \tilde{V}_{\mathbf{n}}/\hbar}. \label{eq:Sommerfeld_Leading_Order}
\end{equation}
Evaluating the sum in Eq.~(\ref{eq:Sommerfeld_Leading_Order}) can only be done exactly for harmonic systems. However, a good approximation can typically be made by considering anharmonic corrections to the transition state free energy. At order $\hbar$ (VPT2) one obtains
\begin{equation}
    kZ_r \approx \frac{e^{-\tau V^\ddagger/\hbar}}{2\pi\tau}\prod_{k=1}^{F-1}\frac{1}{2\sinh(
    \tau\omega_k/2)}e^{-\tau\langle{V}_{\mathbf{n},2}\rangle\hbar}, \label{eq:Sommerfeld_Leading_Order_VPT2}
\end{equation}
where $\langle{V}_{\mathbf{n},2}\rangle=\gamma_0+\zeta(\tau)+\xi(\tau)$ (given in the previous section). Note that, consistent with this, in the absence of rotations the cumulant resummed VPT2 approximation to the reactant partition function is 
\begin{equation}
    Z_r\approx\frac{1}{\Lambda_r}e^{-\tau V_r/\hbar}\left[\prod_{k=1}^{F-1}\frac{1}{2\sinh(\omega_{r,k}\tau/2)}\right]e^{-\tau[\gamma_{0,r}+\zeta_r(\tau)+\xi_r(\tau)]}
\end{equation}
with $\Lambda_r = \sqrt{2\pi\hbar\tau/\mu}$ the thermal de Broglie wavelength for the asymptotic coordinate with mass, $\mu$.
Although this approximation captures quantum effects on the barrier height it does not capture the smooth nature of the transmission probability (i.e.~``tunneling'' and ``above barrier reflection'').

We can incorporate these effects in a consistent manner by noting that  Eq.~(\ref{eq:Sommerfeld_Leading_Order}) is just the leading order contribution to a Sommerfeld expansion of the rate.
Hence, consistent with this approximation, rather than expand in a series in $\hbar$ (as done in the previous sections) we can consider the Sommerfeld expansion of
\begin{equation}
\begin{aligned}
    kZ_r&=\frac{1}{2\pi\hbar} \sum_{\mathbf{n}} \int_{-\infty}^\infty e^{-\tau E/\hbar} \frac{1}{1+e^{\tilde{W}_{\mathbf{n}}(E;\hbar)/(\epsilon\hbar)}} \mathrm{d}E\\
    &\sim \sum_{\mathbf{n}}\frac{e^{-\tau \tilde{V}_{\mathbf{n}}/\hbar}}{2\pi\tau}\left(1+\epsilon^2\left[\frac{\tau^2\tilde{\omega}_{\mathbf{n}}^2}{24}-\frac{\tau\hbar\tilde{W}''_{\mathbf{n}}(\tilde{V}_{\mathbf{n}};\hbar)\tilde{\omega}_{\mathbf{n}}^3}{48\pi}\right]+\mathcal{O}(\epsilon^4)\right)
    \end{aligned}
\end{equation}
as $\epsilon\to0$, where $\tilde{\omega}_{\mathbf{n}}=2\pi/\tilde{\tau}_{\mathbf{n}}(\tilde{V}_{\mathbf{n}};\hbar)$ is the generalised quantum barrier frequency.
Again, the leading order ($\epsilon^0$) term is precisely the ``tunneling free'' rate given by Eq.~(\ref{eq:Sommerfeld_Leading_Order}). The $\epsilon^2$ term corresponds to a generalised multidimensional and anharmonic version of Wigner's harmonic tunneling factor. Expanding the correction factor up to first order in the anharmonicity  (i.e.~order $\hbar$, equivalent to VPT2) we obtain
\begin{equation}
\begin{aligned}
    kZ_r&\sim\sum_{\mathbf{n}}\frac{e^{-\tau \tilde{V}_{\mathbf{n}}/\hbar}}{2\pi\tau}\left(1+\epsilon^2\left[\frac{\tau^2({\omega}^2+2\hbar\omega_{\mathbf{n},1}\omega)}{24}+\frac{\tau\hbar\chi_{F\!F}}{12}\right]\right)
    \end{aligned}
\end{equation}
where $\omega_{\mathbf{n},1}=-\sum_{k=1}^{F-1}\chi_{kF}(n_k+\tfrac{1}{2})$.
We see that the leading order $\hbar^0$ contribution to the $\epsilon^2$ tunneling correction is precisely the harmonic Wigner approximation.
Evaluating the sums at order $\hbar$ (VPT2 level) and performing a simple cumulant resummation of the barrier frequency, we obtain the multidimensional anharmonic Wigner tunneling correction that is consistent with Eq.~(\ref{eq:Sommerfeld_Leading_Order_VPT2}) 
\begin{equation}
    \kappa_{\text{m-anh-tun} }=\left(1+\left[\frac{\tau^2{\omega}^2e^{2\hbar\langle\omega_{\mathbf{n},1}\rangle/\omega}}{24}+\frac{\tau\hbar\chi_{F\!F}}{12}\right]\right)
\end{equation}
where $\langle\omega_{\mathbf{n},1}\rangle = -\sum_{k=1}^{F-1}\chi_{kF}\tfrac{1}{2}\coth(\omega_k\tau/2)$.

We again note the strong connection between this result and the earlier work on anharmonic corrections to the Wigner tunneling correction.\cite{Pollak2022hbarSquared_Correction,Pollak2024hbar4_expansion,Aieta2024AboveBarrierReflection} The key difference is that, whereas Ref.~\citenum{Pollak2022hbarSquared_Correction} included all quantum corrections to the rate in the correction factor, here we have separated the tunneling and zero-point energy contributions to the rate. [In one dimension,  the result obtained here can be related to the fixed $\beta$, $\hbar\to0$ expression of Ref.~\citenum{Pollak2022hbarSquared_Correction}, by replacing $\tau$ with $\beta\hbar$ and expanding the shifted barrier height.] While the present choice to consider quantum changes to the barrier height as zero-point energy contributions rather than tunneling contributions might seem strange in one dimension, this choice is more consistent with the usual practice of quantizing vibrational degrees of freedom in the activated complex.\cite{FernandezRamos2007review} In multiple dimensions following the fixed $\beta$ with $\hbar\to0$ analysis of Refs.~\citenum{Pollak2022hbarSquared_Correction} and \citenum{Pollak2024hbar4_expansion} corresponds to a qualitatively different result, i.e.~perturbative quantum corrections to the fully anharmonic classical rate, in contrast to the present ``tunneling'' correction to the anharmonic quantum rate.

\section{Conclusion and future work}\label{sec:Conclusion}
This paper has introduced a semiclassical framework within which approximate methods for the calculation of reaction rate constants can be derived. 
The formalism rests on a conjectured connection between the ``exact'' action for barrier transmission and the ``exact'' instanton contribution to the trace of the Green's function.\cite{Ture2025ExactWKB,ChaosBookAppendixOnGreensFunction} 
This conjecture can be understood as an alternative to the SCTST ansatz\cite{Miller1977SCTST} that avoids the explicit reference to the concept of ``good'' action variables at the transition state.
We have demonstrated the validity of the conjecture in a number of important cases, and
have shown how the resulting formalism encompasses both instanton theory\cite{Miller1975semiclassical} and VPT2-SCTST.\cite{Miller1990SCTST}
However, the key advantage of the present formalism is that it opens up the opportunity for the development of new methods and the systematic improvement of existing techniques. 

For the calculation of thermal rates we have shown that the present theory can be used to derive the first-order corrections to the rate both above the crossover temperature (i.e.~the parabolic barrier/sphaleron rate) and below the crossover temperature (i.e.~the instanton rate). 
A natural next step is to combine these results with the recently developed uniform asymptotic series method for extending instanton theory to arbitrary temperatures.\cite{Lawrence2024crossover}
We note that the first-order correction to instanton theory derived here must (by the uniqueness of asymptotic power series) be formally equivalent  to the  perturbatively corrected ring-polymer instanton rate expression (RPI+PC) derived using the flux-flux correlation function formalism.\cite{Lawrence2023RPI+PC,Dusek2025RPI+PC}
Confirmation of this equivalence will provide further support to the conjecture.
The present approach also offers advantages over the flux-flux formulation. In particular, it provides a natural route to the derivation of a dividing surface free version of RPI+PC, 
and hence can more easily be generalised to treat systems with rotational degrees of freedom.\cite{InstReview}
Future work will, therefore, look to develop the present formalism within the thermal ring-polymer instanton framework\cite{Andersson2009Hmethane,Richardson2009RPInst,InstReview}  to be able to model chemical reaction rates in full dimensionality at any temperature with the inclusion of anharmonicity and tunneling.

As the formalism presented here is fundamentally microcanonical it provides a natural starting point for the development of new instanton based methods for calculating the cumulative reaction probability.
Unfortunately, although the formalism is formally exact,  a perturbative expansion of $\tilde{W}_{\mathbf{n}}(E;\hbar)$ results in a microcanonical theory that is not size extensive.
This difficulty is not new,\cite{Chapman1975classical} and previous work on microcanonical instanton theory has suggested various practical approaches to overcome the issue. 
\cite{Chapman1975rates,Faraday,DoSTMI,JoeFaraday} 
One may hope that the clarity of the present framework will enable the development of new ideas in this direction.
Of particular significance in this area is the recent suggestion to develop a microcanonical method that combines information from VPT2-SCTST and instanton calculations.\cite{Upadhyayula2024hbar2corrections,Upadhyayula2025CollinearH+H2} 
Building on these ideas, this paper will be followed by an exploration of one such technique that leverages insight gained from the present formalism, such as using the one-dimensional $W_2(E)$, to give an improved method that is more accurate than microcanonical instanton theory or VPT2-SCTST on their own.

The present formalism also provides a natural starting point for the derivation of more approximate theories that only require calculations at the transition state, akin to VPT2-SCTST. 
Here, we have used the formalism to emphasise the precise information one obtains about $\tilde{W}_{\mathbf{n}}(E;\hbar)$ from VPT2-SCTST and highlight the resummation implicit in the standard application of the method.
We have also shown how this information can be combined with a simple Sommerfeld expansion of the exact expression for the thermal rate to give a simple multidimensional anharmonic transition state theory and a corresponding anharmonic generalisation of Wigner's famous tunneling correction factor.\cite{Wigner1932parabolic} 
Of course, both the standard VPT2-SCTST resummation and the Wigner tunneling correction are not accurate in deep tunneling (i.e.~below the crossover temperature).
Future work will, therefore, build on the
 present formalism to develop new methods that are more accurate in deep tunneling. For example, following the work of Wagner,\cite{Wagner2013SCTST} alternative resummations will be considered that combine VPT2-SCTST with information about reactant and product energies effectively generalising  the standard Eckart tunneling correction.\cite{Miller1979unimolecular}

In addition to these practical developments there are also a number of interesting areas for further theoretical investigation.
While the present derivation was entirely heuristic, motivated by inspection of the form of the Green's function, it should be possible to derive the present result from first principles.\cite{Richardson2016FirstPrinciples,InstReview}
One would hope that this first principles derivation would provide additional insight into both the imaginary free-energy (im-F) derivation of instanton theory\cite{Coleman1977ImF}  and the connection of SCTST to the Siegert eigenvalues, emphasised by Seideman and Miller.\cite{Seideman1991SCTST_Siegert_eigvals}
Such a first principles derivation would also be useful in considering systems exhibiting interesting deviations from standard transition state theory assumptions, such as reactions with two or more \emph{interacting} reaction paths. 
In the present work, we have 
focussed for simplicity on systems in the absence of rotational degrees of freedom; future work will look to extend the framework  to treat rotations following e.g.~Ref.~\citenum{Hernandez1993SCTST_rotations}. 

Although we have focussed on molecular scattering on Born-Oppenheimer potential energy surfaces, there are multiple avenues for extension of the formalism outside of this domain.
First, the similarity to the imaginary free-energy (im-F) formulation of instanton theory developed for studying the decay of metastable states\cite{Coleman1977ImF,Callan1977ImF}  suggests a potential route to modelling state-specific decay rates at low energy.\cite{Garbrecht2025ExcitedStateInstantonImF} 
Second, instanton theory has already been generalised and applied with great success to electronically non-adiabatic reactions.\cite{inverted,PhilTransA,thiophosgene,nitrene,4thorder,Fang2023ConicalIntersections,Ansari2024Oxygen,Richardson2024NonAdTunneling,Zarotiadis2025NASCI}
While early work in this area made use of the im-F formalism,\cite{Cao1997nonadiabatic} recent progress has focussed primarily on the flux-correlation formalism;\cite{inverted,PhilTransA,thiophosgene,nitrene,4thorder,Fang2023ConicalIntersections,Ansari2024Oxygen,Richardson2024NonAdTunneling,Zarotiadis2025NASCI} it will be interesting to consider how the ideas presented here can be generalised to such systems and if they can shed light on the deficiencies found in the im-F approach.\cite{Zarotiadis2025NAimF}
Finally, semiclassical techniques such as instanton theory and the Green's function of Eq.~(\ref{eq:Greens_Fn}) have counterparts in the study of stochastic processes,\cite{ChaosBook,Heller2024Instanton} and hence the present formalism may well have counterparts in this alternative context.

In summary, the framework introduced here forms the basis for developing a rigorous and systematically improvable hierarchy of semiclassical reaction rate theories.
Such techniques have the potential to become a serious competitor to the dominant paradigm of variational transition state theory with small and large curvature tunneling corrections in the study of gas-phase chemistry.\cite{FernandezRamos2007review} 
The rigorous and controlled nature of the approximations used in these theories, along with their simple implementation, makes them well suited for integration into automated workflows.
However, one of the key remaining challenges to making this semiclassical framework entirely universal is the ability to treat  hindered rotors and other ``floppy'' low-frequency modes. In the short term these effects can be captured by using suitable correction factors computed with existing methods, e.g.~those available in the MSTor package.\cite{Zheng2012MSTor} However, in the longer term, even these effects can be brought into the domain of rigorous semiclassics using the techniques of uniform asymptotics.\cite{RWong1989UniformAsymptotics}

\section*{Data Availability Statement}
The data that support the findings of this study are available from the corresponding author upon reasonable request.

\section*{Acknowledgements}
This work was supported by an Independent Postdoctoral Fellowship at the Simons Center for Computational Physical Chemistry, under a grant from the Simons Foundation (839534, MT).

\section*{Appendices} \setcounter{section}{0}
\subsection{Eigenreaction probabilities and the scattering matrix}\renewcommand{\theequation}{A\arabic{equation}}
\setcounter{equation}{0} \label{app:Scattering}
The scattering matrix element $S_{\beta \alpha}(E)$ gives the probability amplitude that an incoming state with unit flux incident on the reaction barrier in the infinite past in channel $\alpha$ ends up as an outgoing state with unit flux in channel $\beta$ in the infinite future.\cite{TaylorScatteringBook} In the standard scattering basis, these ``channels'' are chosen to correspond to the individual (energetically accessible) vibrational states of the reactant and product, i.e.~$\alpha$ is an index that labels all energetically accessible $\mathbf{n}_r$ and $\mathbf{n}_p$.
It follows that if $\alpha$ corresponds to state $\mathbf{n}_r$ and $\beta$ corresponds to state $\mathbf{n}_p$ then $|S_{\beta \alpha}(E)|^2$ is the reaction probability that the system transmits from state $\mathbf{n}_r$ of the reactants to   state $\mathbf{n}_p$ of the products.
Note that the scattering matrix also encapsulates non-reactive events, where both $\alpha$ and $\beta$ correspond to states of the reactants or products.

Therefore, if $\alpha$ labels the incoming channel  $\mathbf{n}_r$, then summing over all outgoing product channels recovers 
\begin{subequations}
    \begin{equation}
    P_{\mathbf{n}_r}(E)=\sum_{\beta\in p}|S_{\beta\alpha}(E)|^2
\end{equation}
or for $\alpha$ corresponding to $\mathbf{n}_p$ summing over outgoing reactant channels gives
\begin{equation}
    P_{\mathbf{n}_p}(E)=\sum_{\beta\in r}|S_{\beta\alpha}(E)|^2
\end{equation}
\end{subequations}
and, hence, $N(E) = \sum_{\alpha\in r,\beta\in p}|S_{\beta\alpha}(E)|^2$.

With this in hand we can now more properly illuminate the kind of basis transformation that we only alluded to in the main text. In particular, one can choose to perform a basis rotation between the channels. An especially suggestive example of such a basis transformation is one where the reactant incoming states transmit to only one product channel and reflect only to themselves, i.e.~a set of eigenchannels with corresponding eigenreaction probabilities.\cite{Mello1988ScatteringMatrix,Manthe1993EigenreactionProbabilities} Defining the scattering matrix in the original channel basis as
\begin{equation}
    \mathbf{S}(E) = \begin{bmatrix} \mathbf{r} & \mathbf{t}' \\
    \mathbf{t} & \mathbf{r}'\end{bmatrix},
\end{equation}
where $\mathbf{r}$ and $\mathbf{t}$  are the blocks corresponding to reflection and transmission for states incident from the reactants, and $\mathbf{r}'$ and $\mathbf{t}'$ are the blocks corresponding to reflection and transmission for states incident from the products, the eigenchannel basis transformation is then achieved by block diagonalizing the scattering matrix,
\begin{equation}
    \mathbf{S}(E) = \begin{bmatrix} \mathbf{V}^* & \mathbf{0} \\
    \mathbf{0} & \mathbf{U}\end{bmatrix} \begin{bmatrix} -(\mathbf{1}-\mathbf{T})^{1/2} & \mathbf{T}^{1/2} \\
    \mathbf{T}^{1/2} & (\mathbf{1}-\mathbf{T})^{1/2} \end{bmatrix} \begin{bmatrix} \mathbf{V}^\dagger & \mathbf{0} \\
    \mathbf{0} & \mathbf{U}^{\rm T}\end{bmatrix}
\end{equation}
where the matrices $\mathbf{U}$ and $\mathbf{V}$ are defined by a singular value decomposition of the transmission matrix,
\begin{equation}
    \mathbf{t}=\mathbf{U} \mathbf{T}^{1/2} \mathbf{V}^{\dagger},
\end{equation}
and $\mathbf{T}$ is a diagonal matrix of eigenchannel transmission probabilities (eigenreaction probabilities),\cite{Manthe1993EigenreactionProbabilities} such that $N(E)=\sum_\alpha T_{\alpha}(E)$. That this recovers the original scattering matrix can be shown by making use of the unitarity $\mathbf{S}^\dagger(E)\mathbf{S}(E)=1$, and symmetry, $\mathbf{S}(E)=\mathbf{S}^{\rm T}(E)$ of the scattering matrix.\cite{Mello1988ScatteringMatrix}
Hence, transforming the incoming basis of states by $\mathbf{W}(E)=\text{diag}(\mathbf{V}^\dagger(E),\mathbf{U}^{\rm T}(E))$ and the outgoing by $\mathbf{W}^*(E)$ 
 corresponds to an \emph{energy dependent} basis transformation that effectively turns the multidimensional scattering system into a set of uncoupled one-dimensional scattering problems. 

\subsection{Fourth order correction to the action}\renewcommand{\theequation}{B\arabic{equation}}\setcounter{equation}{0} \label{app:W_4}
A simple asymptotic analysis (expanding everything in $\hbar$) following the exact WKB\cite{Ture2025ExactWKB} approach outlined in Sec.~\ref{sec:One_dimension} results in the following expression for the $\hbar^4$ expansion coefficient of the exact action for the instanton
\begin{equation}\begin{aligned}
    W_4(E)&=\frac{1}{128} \oint \frac{-4[V''(x)]^2}{[p_0(x)]^7}\mathrm{d}x\\
    &+\frac{1}{128} \oint\frac{380[V'(x)]^2V''(x)}{[p_0(x)]^9}\mathrm{d}x\\&-\frac{1}{128} \oint\frac{1105[V'(x)]^4}{[p_0(x)]^{11}} \mathrm{d}x \label{eq:W_4_1D}
    \end{aligned}
\end{equation}
where $x$ is the mass-weighted coordinate and the classical momentum is
\begin{equation}
    p_0(x)= \sqrt{2(V(x)-E)}.
\end{equation}

\subsection{Eckart barrier exact action}\renewcommand{\theequation}{C\arabic{equation}}
\setcounter{equation}{0} \label{app:Eckart_Action}
For the asymmetric Eckart barrier the exact transmission probability is given by\cite{Eckart,Miller1979unimolecular,Upadhyayula2024hbar2corrections,BellBook} 
\begin{equation}
    P(E) = \frac{\cosh(\frac{A}{\hbar}\left(\varphi_1(E)+\varphi_2(E)\right))-\cosh(\frac{A}{\hbar}\left(\varphi_1(E)-\varphi_2(E)\right))}{\cosh(\frac{A}{\hbar}\left(\varphi_1(E)+\varphi_2(E)\right))+\cosh(\frac{A}{\hbar}\varphi_3(\hbar))}
\end{equation}
where we have defined
\begin{subequations}
\begin{equation}
    A = \frac{4\pi}{1+\sqrt{V_2/V_1}}\sqrt{\frac{V_1 V_2}{\omega^2}}
\end{equation}
\vspace{-0.6cm}
    \begin{equation}
        \varphi_1(E) = \sqrt{\frac{E}{V_1}}
    \end{equation}
    \vspace{-0.6cm}
    \begin{equation}
        \varphi_2(E) = \sqrt{\frac{E}{V_1}-1+\frac{V_2}{V_1}}
    \end{equation}
    \vspace{-0.6cm}
    \begin{equation}
        \varphi_3(\hbar) = \left(1+\sqrt{\frac{V_2}{V_1}}\right)\sqrt{1-\frac{\hbar^2\omega^2}{16V_1V_2}}.
    \end{equation}
\end{subequations}
Hence, we have that
\begin{equation}
    \tilde{W}(E;\hbar) = \hbar \ln(\frac{\cosh(\frac{A\left(\varphi_1(E)-\varphi_2(E)\right)}{\hbar})+\cosh(\frac{A\varphi_3(\hbar)}{\hbar})}{\cosh(\frac{A\left(\varphi_1(E)+\varphi_2(E)\right)}{\hbar})-\cosh(\frac{A\left(\varphi_1(E)-\varphi_2(E)\right)}{\hbar})}).
\end{equation}
Writing the hyperbolic cosines in terms of exponentials and noting that when $V_2>V_1$ it follows $\varphi_2(E)>\varphi_1(E)$, and that $\varphi_3(\hbar)>(\varphi_2(E)-\varphi_1(E))$ (as $\hbar\to 0$) allows us to identify the dominant terms
\begin{equation}
\begin{aligned}
    \tilde{W}(E;\hbar) =& \hbar \ln(e^{\frac{A(\varphi_1-\varphi_2)}{\hbar}}+e^{-\frac{A(\varphi_1-\varphi_2)}{\hbar}}+e^{\frac{A\varphi_3}{\hbar}}+e^{-\frac{A\varphi_3}{\hbar}}) \\
    &-\hbar \ln(e^{\frac{A(\varphi_1+\varphi_2)}{\hbar}}+e^{-\frac{A(\varphi_1+\varphi_2)}{\hbar}}-e^{\frac{A(\varphi_1-\varphi_2)}{\hbar}}-e^{-\frac{A(\varphi_1-\varphi_2)}{\hbar}}) \\
    =&A\varphi_3(\hbar)+\hbar\ln(1+e^{\frac{A(\varphi_1-\varphi_2-\varphi_3)}{\hbar}}+e^{-\frac{A(\varphi_1-\varphi_2+\varphi_3)}{\hbar}}+e^{-\frac{2A\varphi_3}{\hbar}}) \\
    &-A(\varphi_1(E)+\varphi_2(E)) \\ & - \hbar\ln(1+e^{-\frac{2A(\varphi_1+\varphi_2)}{\hbar}}-e^{-\frac{2A\varphi_2}{\hbar}}-e^{-\frac{2A\varphi_1}{\hbar}})
    \end{aligned}
\end{equation}
Hence, expanding the logarithms gives
\begin{equation}
    \tilde{W}(E;\hbar)\sim A(\phi_3(\hbar)-\phi_1(E)-\phi_2(E))
\end{equation}
with correction terms that are exponentially small as $\hbar\to0$. Expanding $\phi_3(\hbar)$ in a series in $\hbar$ we then obtain explicit expressions for the $W_{2n}(E)$ given in Eq.~(\ref{eq:Eckart_W2n}).

\subsection{VPT2 anharmonicity constants for minima}\renewcommand{\theequation}{D\arabic{equation}}
\setcounter{equation}{0} \label{app:VPT2_constants}
Here for completeness we give the ($\hbar$ independent) VPT2 constants in our notation. In keeping with the original work of Miller \emph{et al.}\cite{Miller1990SCTST} we define the cubic and quartic force constants as
\begin{subequations}
\begin{equation}
    f_{klm}=\frac{\partial^3 V}{\partial Q_k\partial Q_l\partial Q_m}
\end{equation}
\begin{equation}
    f_{klmn}=\frac{\partial^4 V}{\partial Q_k\partial Q_l\partial Q_m\partial Q_n},
\end{equation}
\end{subequations}
where $\{Q_k\}$ are the mass-weighted normal mode coordinates at the stationary point.

For the reactants we have, the diagonal anharmonicity constants are given by
\begin{equation}
    \chi_{kk} = \frac{1}{16\omega_k^2}\left( f_{kkkk} - \sum_{l=1}^{F_r} \frac{f_{kkl}^2(8\omega_k^2-3\omega_l^2)}{\omega_l^2(4\omega_k^2-\omega_l^2)}\right)
\end{equation}
where in the case of escape from a metastable well $F_r=F$ and in the case that there is a single zero mode $F_r=F-1$. The off-diagonal anharmonicity constants are given by
\begin{equation}
\begin{aligned}
    \chi_{kl}&= \frac{1}{4\omega_k\omega_l}\Bigg(f_{kkll}-\sum_{m=1}^{F_r}\frac{f_{kkm}f_{llm}}{\omega_m^2}\\&+\sum_{m=1}^{F_r}\frac{2f_{klm}^2(\omega_k^2+\omega_l^2-\omega_m^2)}{[(\omega_k+\omega_l)^2-\omega_m^2][(\omega_k-\omega_l)^2-\omega_m^2]}\Bigg).
    \end{aligned}
\end{equation}
Finally the zero-point constant is given by\cite{Truhlar1991ZPE_VPT2,Hernandez1993SCTST}
\begin{equation}
\begin{aligned}
    \gamma_{0,r} &= \frac{1}{64}\sum_{k=1}^{F_r}\frac{f_{kkkk}}{\omega_k^2}-\frac{7}{576}\sum_{k=1}^{F_r}\frac{f_{kkk}^2}{\omega_k^4}+\frac{3}{64}\sum_{k\neq l}^{F_r} \frac{f_{kll}^2}{(4\omega_l^2-\omega_k^2)\omega_l^2} \\
    &-\frac{1}{4}\sum_{k<l<m}^{F_r}\frac{f_{klm}^2}{[(\omega_k+\omega_l)^2-\omega_m^2][(\omega_k-\omega_l)^2-\omega_m^2]}.
    \end{aligned}
\end{equation}

\subsection{SCTST/VPT2 constants for transition states}\renewcommand{\theequation}{E\arabic{equation}}
\setcounter{equation}{0} \label{app:SCTST/VPT2_constants}
Here we give the SCTST/VPT2 constants\cite{Miller1990SCTST} in the present notation. 
For $k=1,\dots,F-1$ we have the diagonal anharmonic constants are given by
\begin{equation}
    \chi_{kk} = \frac{1}{16\omega_k^2}\left(f_{kkkk}+\frac{f_{kkF}^2(8\omega_k^2+3\omega^2)}{\omega^2(4\omega_k^2+\omega^2)}-\sum_{l=1}^{F-1}\frac{f_{kkl}^2(8\omega_k^2-3\omega_l^2)}{\omega_l^2(4\omega_k^2-\omega_l^2)}\right) 
\end{equation}
and for the unstable coordinate we have
\begin{equation}
    \chi_{F\!F}=\frac{1}{16\omega^2}\left(-f_{F\!F\!F\!F}-\frac{5}{3}\frac{f^2_{F\!F\!F}}{\omega^2}+\sum_{l=1}^{F-1}\frac{f_{F\!Fl}^2(8\omega^2+3\omega_l^2)}{\omega_l^2(4\omega^2+\omega_l^2)}\right).
\end{equation}
For $k,l=1,\dots, F-1$ the off-diagonal anharmonic constants are given by 
\begin{equation}
\begin{aligned}
    \chi_{kl} = &\frac{1}{4\omega_{k}\omega_l}\Bigg[f_{kkll}+\frac{f_{kkF}f_{llF}}{\omega^2}\\+&\frac{2f_{klF}^2(\omega_k^2+\omega_l^2+\omega^2)}{[(\omega_k+\omega_l)^2+\omega^2][(\omega_k-\omega_l)^2+\omega^2]}\\-&\sum_{m=1}^{F-1}\left(\frac{f_{kkm}f_{llm}}{\omega_m^2}-\frac{2f_{klm}^2(\omega_k^2+\omega_l^2-\omega_m^2)}{[(\omega_k+\omega_l)^2-\omega_m^2][(\omega_k-\omega_l)^2-\omega_m^2]}\right)\Bigg]
    \end{aligned}
\end{equation}
and for $k=1,\dots,F-1$
\begin{equation}
\begin{aligned}
    \chi_{kF} &= \frac{1}{4\omega_{k}\omega}\Bigg[f_{kkF\!F}+\frac{f_{kkF}f_{F\!F\!F}}{\omega^2}+\frac{2f_{kF\!F}^2}{\omega_k^2+4\omega^2}\\-&\sum_{m=1}^{F-1}\left(\frac{f_{kkm}f_{F\!Fm}}{\omega_m^2}-\frac{2f_{kFm}^2(\omega_k^2-\omega^2-\omega_m^2)}{[(\omega_k+\omega_m)^2+\omega^2][(\omega_k-\omega_m)^2+\omega^2]}\right)\Bigg].
    \end{aligned}
\end{equation}
Finally the zero-point constant is\cite{Truhlar1991ZPE_VPT2,Hernandez1993SCTST}
\begin{equation}
\begin{aligned}
    \gamma_0 &= \frac{1}{64}\sum_{k=1}^{F-1}\frac{f_{kkkk}}{\omega_k^2}-\frac{7}{576}\sum_{k=1}^{F-1}\frac{f_{kkk}^2}{\omega_k^4}+\frac{3}{64}\sum_{k\neq l}^{F-1} \frac{f_{kll}^2}{(4\omega_l^2-\omega_k^2)\omega_l^2} \\
    &-\frac{1}{4}\sum_{k<l<m}^{F-1}\frac{f_{klm}^2}{[(\omega_k+\omega_l)^2-\omega_m^2][(\omega_k-\omega_l)^2-\omega_m^2]}\\
    &-\frac{f_{F\!F\!F\!F}}{64\omega^2}-\frac{7f_{F\!F\!F}^2}{576\omega^4} +\frac{3}{64}\sum_{k=1}^{F-1}\frac{f_{kF\!F}^2}{(4\omega^2+\omega_k^2)\omega^2}\\
    &+\frac{3}{64}\sum_{k=1}^{F-1}\frac{f_{kkF}^2}{(4\omega_k^2+\omega^2)\omega_k^2} \\
    &-\frac{1}{4}\sum_{k<l}^{F-1} \frac{f_{klF}^2}{[(\omega_k+\omega_l)^2+\omega^2][(\omega_k-\omega_l)^2+\omega^2]}.
    \end{aligned}
\end{equation}

\subsection{SCTST/VPT2 equating coefficients}\label{app:W_from_VPT2} \renewcommand{\theequation}{F\arabic{equation}}
\setcounter{equation}{0} 
Here we determine what information is obtained about $\tilde{W}_{\mathbf{n}}(E;\hbar)$ from VPT2. 
To achieve this we begin by expanding the coefficients of Eq.~(\ref{eq:Action_expansion_around_V_tilde}) in powers of $\hbar$ up to second order to give
\begin{equation}
\begin{aligned}
    \tilde{W}_{\mathbf{n}}^{(m)}(\tilde{V}_{\mathbf{n}};\hbar) =& W_{0}^{(m)}(V^\ddagger) + W_{0}^{(m+1)}(V^\ddagger)\left(V_{\mathbf{n},1}\hbar+V_{\mathbf{n},2}\hbar^2+\dots\right)\\ &
    +\frac{1}{2}W_{0}^{(m+2)}(V^\ddagger)(V_{\mathbf{n},1}\hbar+\dots)^2+ \dots \\
    & + W_{\mathbf{n},1}^{(m)}(V^\ddagger)\hbar  + W_{\mathbf{n},1}^{(m+1)}(V^\ddagger)\hbar\left(V_{\mathbf{n},1}\hbar+\dots\right)+\dots\\
    &+W_{\mathbf{n},2}^{(m)}(V^\ddagger)\hbar^2 + \dots
    \end{aligned}
\end{equation}
where in keeping with the notation $\tilde{V}_{\mathbf{n}}=a_{\mathbf{n},0}$ we have also defined the expansion coefficients of $\tilde{V}_{\mathbf{n}}$ as  $a_{\mathbf{n},0,m}=V_{\mathbf{n},m}$.
The next step is to expand the explicit expressions for the coefficients in terms of the $a_{\mathbf{n},\nu}(\hbar)$ from Eq.~(\ref{eq:a_n_nu_expansion}) to give
\begin{subequations}
\begin{equation}
    \tilde{W}_{\mathbf{n}}^{(0)}(\tilde{V}_{\mathbf{n}}(\hbar);\hbar) = 0
\end{equation}
\begin{equation}
    \tilde{W}_{\mathbf{n}}^{(1)}(\tilde{V}_{\mathbf{n}}(\hbar);\hbar) = \frac{1}{a_{\mathbf{n},1}(\hbar)} = \frac{1}{a_{\mathbf{n},1,0}} - \hbar\frac{a_{\mathbf{n},1,1}} {a_{\mathbf{n},1,0}^2} + \mathcal{O}(\hbar^2)
\end{equation}
\begin{equation}
    \tilde{W}_{\mathbf{n}}^{(2)}(\tilde{V}_{\mathbf{n}}(\hbar);\hbar) = -\frac{2a_{\mathbf{n},2}(\hbar)}{a_{\mathbf{n},1}^3(\hbar)} = -\frac{2a_{\mathbf{n},2,0}}{a_{\mathbf{n},1,0}^3} + \mathcal{O}(\hbar)
\end{equation}
\end{subequations}
Where we have only retained terms up to the order to which VPT2 is exact. Now we can equate coefficients to obtain the following relations, first from $\tilde{W}_{\mathbf{n}}^{(0)}(\tilde{V}_{\mathbf{n}}(\hbar);\hbar)$ we have
\begin{subequations}
    \begin{equation}
       W_{0}(V^\ddagger) = 0
\end{equation}
    \begin{equation}
       W_{0}^{(1)}(V^\ddagger)V_{\mathbf{n},1} + W_{\mathbf{n},1}^{(0)}(V^\ddagger)  = 0
\end{equation}
    \begin{equation}
      W_{0}^{(1)}(V^\ddagger)V_{\mathbf{n},2}+ \frac{1}{2}W_{0}^{(2)}(V^\ddagger)V_{\mathbf{n},1}^2 + W_{\mathbf{n},1}^{(1)}(V^\ddagger)V_{\mathbf{n},1}  + W_{\mathbf{n},2}^{(0)}(V^\ddagger)= 0
\end{equation}
\end{subequations}
and from $\tilde{W}_{\mathbf{n}}^{(1)}(\tilde{V}_{\mathbf{n}}(\hbar);\hbar)$ we have
\begin{subequations}
        \begin{equation}
       W_{0}^{(1)}(V^\ddagger) = \frac{1}{a_{\mathbf{n},1,0}}
\end{equation}
        \begin{equation}
       W_{0}^{(2)}(V^\ddagger)V_{\mathbf{n},1} + W_{\mathbf{n},1}^{(1)}(V^\ddagger) = -\frac{a_{\mathbf{n},1,1}}{a_{\mathbf{n},1,0}^2}
\end{equation}
\end{subequations}
and finally from $\tilde{W}_{\mathbf{n}}^{(2)}(\tilde{V}_{\mathbf{n}}(\hbar);\hbar)$ we have
\begin{equation}
    W_{0}^{(2)}(V^\ddagger) = -\frac{2a_{\mathbf{n},2,0}}{a_{\mathbf{n},1,0}^3}.
\end{equation}
These can then be solved to give
\begin{equation}
        W_{\mathbf{n},1}^{(1)}(V^\ddagger) = -\frac{a_{\mathbf{n},1,1}}{a_{\mathbf{n},1,0}^2} + \frac{2a_{\mathbf{n},2,0}}{a_{\mathbf{n},1,0}^3}V_{\mathbf{n},1}
\end{equation}
\begin{equation}
         W_{\mathbf{n},1}^{(0)}(V^\ddagger)  = -\frac{V_{\mathbf{n},1}}{a_{\mathbf{n},1,0}}
\end{equation}
    \begin{equation}
           W_{\mathbf{n},2}^{(0)}(V^\ddagger)= \frac{a_{\mathbf{n},1,1}}{a_{\mathbf{n},1,0}^2} V_{\mathbf{n},1}-\frac{V_{\mathbf{n},2}}{a_{\mathbf{n},1,0}} -\frac{a_{\mathbf{n},2,0}}{a_{\mathbf{n},1,0}^3}V_{\mathbf{n},1}^2 .
\end{equation}

Hence, we see that VPT2 determines all of $W_{0}^{}(V^\ddagger)$, $W_{0}^{(1)}(V^\ddagger)$, $W_{0}^{(2)}(V^\ddagger)$, $W_{\mathbf{n},1}^{}(V^\ddagger)$, $W_{\mathbf{n},1}^{(1)}(V^\ddagger)$, and $W_{\mathbf{n},2}^{}(V^\ddagger)$, and only these constants. The SCTST assumption that all other $a_{\mathbf{n},\nu,m}=0$ can be viewed as an approximate resummation scheme. 
Finally, using the identities
\begin{subequations}
\begin{equation}
    V_{\mathbf{n},1} = \sum_{k=1}^{F-1}  \omega_k\left(n_k+\tfrac{1}{2}\right)
\end{equation}
\begin{equation}
    V_{\mathbf{n},2} = \gamma_0 + \sum_{k'\leq k=1}^{F-1} \chi_{kk'} \left(n_k+\tfrac{1}{2}\right)\left(n_{k'}+\tfrac{1}{2}\right)
\end{equation}
\begin{equation}
    a_{\mathbf{n},1}(\hbar)  =-\frac{\omega}{2\pi} + \hbar \sum_{k=1}^{F-1} \frac{\chi_{kF}}{2\pi} \left(n_k+\tfrac{1}{2}\right) + \dots := -\frac{\tilde{\omega}_{\mathbf{n}}}{2\pi}
\end{equation}
\begin{equation}
    a_{\mathbf{n},2}(\hbar) = -\frac{\chi_{F\!F}}{4\pi^2} + \dots := -\frac{\tilde{\chi}_{F\!F,\mathbf{n}}}{4\pi^2}
\end{equation}
\end{subequations}
we obtain the result given in the main text.

\subsection{Derivation of thermal instanton rate theory and first-order correction} \renewcommand{\theequation}{G\arabic{equation}}
\setcounter{equation}{0}\label{app:instanton_connection}

Here we give the details of the derivation of the thermal instanton rate theory, as well as the ``multiple bounce'' terms, up to first order in $\hbar$. We begin by defining the integrals up to the effective barrier height in Eq.~(\ref{eq:full_rate_sums}) as
\begin{equation}
    I_{n}(\hbar) = \frac{1}{2\pi\hbar} \sum_{\mathbf{n}}  \int_{-\infty}^{\tilde{V}_\mathbf{n}} e^{-\tau E/\hbar}  e^{-nW_\mathbf{n}(E;\hbar)/\hbar} \, \mathrm{d}E.
\end{equation}
To remove the $\mathbf{n}$ and $\hbar$ dependence from the integration limit, which will allow us to move the sum over $\mathbf{n}$ inside the integral, we can make the variable change $E\to E+\Delta \tilde{V}_{\mathbf{n}}$, where $\Delta\tilde{V}_{\mathbf{n}}=\tilde{V}_{\mathbf{n}}-V^\ddagger$, to give
\begin{equation}
    I_{n}(\hbar) = \frac{1}{2\pi\hbar}   \int_{-\infty}^{V^\ddagger} \sum_{\mathbf{n}}e^{-\tau E/\hbar-\tau\Delta\tilde{V}_{\mathbf{n}}/\hbar}  e^{-nW_\mathbf{n}(E+\Delta\tilde{V}_{\mathbf{n}};\hbar)/\hbar} \, \mathrm{d}E.
\end{equation}
To obtain an asymptotic expansion of this integral as $\hbar\to0$  one then expands the terms appearing in the exponent as series in $\hbar$. 
 First, the effective barrier height can be expanded as
\begin{equation}
    \Delta \tilde{V}_{\mathbf{n}}\sim  \sum_{j=1}^\infty V_{\mathbf{n},j}\hbar^j
\end{equation}
this can then be combined with the definition of the action [Eq.~(\ref{eq:Key_ansatz_2})] to give 
\begin{equation}
\begin{aligned}
        \tilde{W}_{\mathbf{n}}(E+\Delta\tilde{V}_{\mathbf{n}};\hbar) \sim W_0(E) + \hbar \left( W_{\mathbf{n},1}(E) + W_0'(E)V_{\mathbf{n},1} \right) +\mathcal{O}(\hbar^2).
\end{aligned}
\end{equation}
Therefore, at leading order in $\hbar$
\begin{equation}
    I_{n}(\hbar) \sim \frac{1}{2\pi\hbar}   \int_{-\infty}^{V^\ddagger} e^{-\tau E/\hbar-nW_0(E)/\hbar}  Z_{n}(E;\tau) \left[1+\mathcal{O}(\hbar) \right]  \mathrm{d}E \label{eq:I_n_unintegrated_summed}
\end{equation}
where the ``partition function,'' $Z_{n}(E;\tau)$, is defined as
\begin{equation}
    Z_{n}(E;\tau) =  \sum_{\mathbf{n}} e^{-\tau{V}_{\mathbf{n},1}-n\left[W_{\mathbf{n},1}(E)+W_0'(E)V_{\mathbf{n},1}\right]}\label{eq:generalised_partition_function_n}.
\end{equation}
It is then trivial to integrate by steepest descent (see Appendix~\ref{app:steepest_descent_integration}) to obtain the leading order ``instanton'' result. Defining the stationary energy, $E^\star_n$, as $W'_0(E_n^\star)=-\tau/n$, we have that, for $E^\star_n<V^\ddagger$ or equivalently $\tau>2\pi n/\omega$,
\begin{equation}
\begin{aligned}
    I_n(\hbar) &\sim I_{n,0,\mathrm{i}}(\hbar) \\
    I_{n,0,\mathrm{i}}(\hbar) &= \frac{1}{2\pi\hbar} \sqrt{\frac{2\pi\hbar}{nW''_0(E_n^\star)}}e^{-\tau E_n^\star/\hbar-nW_0(E_n^\star)/\hbar} Z_n(E_n^\star;\tau). 
\end{aligned}
\end{equation}
Defining the total instanton action
\begin{equation}
    S_{\!n,{\rm inst}}(\tau) = nW_0(E_n^\star)+\tau E_n^\star
\end{equation}
and noting that at $E_n^\star$ all dependence on $V_{\mathbf{n},1}$ cancels, such that the partition function simplifies to give the usual instanton partition function for $n$ orbits
\begin{equation}
\begin{aligned}
    Z_n(E_n^\star;\tau)&=\sum_{\mathbf{n}} e^{-n W_{\mathbf{n},1}(E_n^\star)} \\&= \sum_{\mathbf{n}} e^{- \sum_k nu_k(E_n^\star) \left(n_k+\tfrac{1}{2}\right)}\\ &=\prod_{k=1}^{F-1}\frac{1}{2\sinh(n u_k(E_n^\star)/2)} = Z_{n,\rm inst}(E_n^\star(\tau))
    \end{aligned}
\end{equation}
we obtain the standard instanton expression
\begin{equation}
\begin{aligned}
    I_{n,0,\mathrm{i}}(\hbar) &=  \sqrt{\frac{-S''_{\!n,\mathrm{inst}}(\tau)}{2\pi\hbar}}e^{-S_{\!n,\mathrm{inst}}(\tau)/\hbar} Z_{n,\rm inst}(\tau)\equiv (k Z_r)_{n,\rm inst,0}. 
\end{aligned}
\end{equation}
For $\tau<\tau_{\rm c}=2\pi/\omega$ all other terms are subdominant to $I_{1,0,{\mathrm{i}}}(\hbar)$ (as $\hbar\to0$) and, hence, as stated in the main text we recover the standard instanton result 
\begin{equation}
    k Z_r \sim (k Z_r)_{1,\rm inst,0} \equiv (k Z_r)_{\rm inst,0}.
\end{equation}

Following the same procedure, we can go beyond the standard leading order instanton approximation and derive higher order asymptotic corrections.  To obtain the first-order correction we thus need to include terms up to $\hbar^2$ in the effective barrier height and exact action, such that we obtain
\begin{equation}
\begin{aligned}
        \tilde{W}_{\mathbf{n}}(E+\Delta\tilde{V}_{\mathbf{n}};\hbar) \sim&   W_0(E) + \hbar \left( W_{\mathbf{n},1}(E) + W_0'(E)V_{\mathbf{n},1} \right) \\
    &+\hbar^2\Bigg( W_{\mathbf{n},2}(E)+ W'_{\mathbf{n},1}(E)V_{\mathbf{n},1} \\&+W_0'(E)V_{\mathbf{n},2} + \frac{W''_0(E)}{2} V_{\mathbf{n},1}^2 \Bigg) + \dots.
\end{aligned}
\end{equation}
Expanding the exponential and again performing the summation over $\mathbf{n}$ we arrive at
\begin{equation}
    I_{n}(\hbar) \sim \frac{1}{2\pi\hbar}   \int_{-\infty}^{V^\ddagger} e^{-\tau E/\hbar-nW_0(E)/\hbar}  Z_{n}(E;\tau) \left[1-\hbar \langle \mu_{\mathbf{n},n}(E)\rangle \right]  \mathrm{d}E
\end{equation}
where
\begin{equation}
    \begin{aligned}
    \mu_{\mathbf{n},n}(E;\tau) = \tau V_{\mathbf{n},2} + n\Bigg(& W_{\mathbf{n},2}(E)+ W'_{\mathbf{n},1}(E)V_{\mathbf{n},1} \\&+W_0'(E)V_{\mathbf{n},2} + \frac{W''_0(E)}{2} V_{\mathbf{n},1}^2 \Bigg) 
    \end{aligned}
\end{equation}
and the expectation value is defined as
\begin{equation}
    \langle \mu_{\mathbf{n},n}(E)\rangle = \frac{\sum_{\mathbf{n}}e^{-\tau{V}_{\mathbf{n},1}-n\left[W_{\mathbf{n},1}(E)+W_0'(E)V_{\mathbf{n},1}\right]} \mu_{\mathbf{n},n}(E)}{Z_{n}(E;\tau)}
\end{equation}
Making use of the standard result given in Appendix~\ref{app:steepest_descent_integration}, we obtain a formal expression for the first-order correction of the form
\begin{equation}
\begin{aligned}
    \!\!\alpha_{n,1,\mathrm{i}} \!&=\! \frac{Z_{n}^{(2)}(E_n^\star;\tau)}{2nZ_n(E_n^\star;\tau)W_0^{(2)}(E_n^\star)}-\frac{Z_n^{(1)}(E_n^\star;\tau)W_0^{(3)}(E_n^\star)}{2nZ_n(E_n^\star;\tau)[W_0^{(2)}(E_n^\star)]^{2}}\\ &-\frac{W_0^{(4)}(E_n^\star)}{8n[W_0^{(2)}(E_n^\star)]^{2}}+\frac{5[W_0^{(3)}(E_n^\star)]^2}{24n[W_0^{(2)}(E_n^\star)]^{3}} -\left\langle \mu_{\mathbf{n},n}(E_n^\star)\right\rangle. \!\!\!\!
    \end{aligned}
\end{equation}
where we can simplify the expectation value to
\begin{equation}
    \mu_{\mathbf{n},n}(E_n^\star) = \ n\Bigg( W_{\mathbf{n},2}(E_n^\star)+ W'_{\mathbf{n},1}(E_n^\star)V_{\mathbf{n},1} + \frac{W''_0(E_n^\star)}{2} V_{\mathbf{n},1}^2 \Bigg). 
\end{equation}
Note that the derivatives, $Z_n^{(\nu)}(E_n^\star;\tau)$, are derivatives of Eq.~(\ref{eq:generalised_partition_function_n}) with respect to energy and hence are \emph{not} the same as derivatives of the instanton partition function $Z_{n,\rm inst}(\tau(E))$. We can simplify by relating the derivatives of $Z_n^{(\nu)}(E_n^\star;\tau)$ to the derivatives of  $Z_{n,\rm inst}(E^\star)$ and cancelling terms to obtain
\begin{equation}
\begin{aligned}
    \!\!\alpha_{n,1,\mathrm{i}} \!&=\! \frac{Z_{n,\rm inst}^{(2)}(E_n^\star)}{2nZ_{n,\rm inst}(E_n^\star)W_0^{(2)}(E_n^\star)}-\frac{Z_{n,\rm inst}^{(1)}(E_n^\star)W_0^{(3)}(E_n^\star)}{2nZ_{n,\rm inst}(E_n^\star)[W_0^{(2)}(E_n^\star)]^{2}}\\ &-\frac{W_0^{(4)}(E_n^\star)}{8n[W_0^{(2)}(E_n^\star)]^{2}}+\frac{5[W_0^{(3)}(E_n^\star)]^2}{24n[W_0^{(2)}(E_n^\star)]^{3}} -n\left\langle W_{\mathbf{n},2}(E_n^\star)\right\rangle. \!\!\!\!
    \end{aligned}
\end{equation}

\subsection{Derivation of the thermal semiclassical rate constant above the crossover temperature and first-order correction to the ``sphaleron'' rate} \renewcommand{\theequation}{H\arabic{equation}}
\setcounter{equation}{0} \label{app:sphaleron_derivation}
In the previous section we have considered the so-called ``instanton'' behaviour when $\tau>2\pi n/\omega$ ($E^\star_n<V^\ddagger$). At high temperatures one must use an alternative asymptotic approximation. As we shall see, at leading order this gives rise to the well-known parabolic barrier approximation to the rate, however, the present theory also allows us to trivially derive the higher order asymptotic corrections to the parabolic barrier rate. As these corrections involve higher derivatives of the potential, it is inappropriate to label them as harmonic or parabolic, we therefore use the term from high energy physics ``sphaleron''.\cite{Klinkhamer1984Sphaleron}

\subsubsection{Leading order contribution: The parabolic barrier rate}
Asymptotic integration of Eq.~(\ref{eq:I_n_unintegrated_summed})   when $E^\star_n>V^\ddagger$, [see Appendix~\ref{app:steepest_descent_integration}], gives the leading order ``sphaleron'' contribution as
\begin{equation}
    I_{n,0,\mathrm{s}}(\hbar) = \frac{1}{2\pi}e^{-\tau V^\ddagger/\hbar} \frac{1}{n\tau_{\rm c} - \tau } Z_{ h}^\ddagger(\tau),
\end{equation}
where we have used the fact that $u_k(V^\ddagger)=\tau_{\rm c} \omega_k$, leading again to significant simplification in the partition function,  $Z_n(V^\ddagger;\tau)=Z_{ h}^\ddagger(\tau)$, with
\begin{equation}
    Z_{h}^\ddagger(\tau) = \prod_{k=1}^{F-1}\frac{1}{2\sinh(
    \tau\omega_k/2)}
\end{equation}
the usual harmonic approximation to the transition state partition function. Above the crossover one must also consider the other integrals in Eq.~(\ref{eq:full_rate_sums})
\begin{equation}
    J_n(\hbar) = \frac{1}{2\pi\hbar} \sum_{\mathbf{n}} \int_{\tilde{V}_\mathbf{n}}^{\infty} e^{-\tau E/\hbar}  e^{nW_\mathbf{n}(E;\hbar)/\hbar} \, \mathrm{d}E.
\end{equation}
These can be similarly manipulated to give
\begin{equation}
    J_{n}(\hbar) \sim \frac{1}{2\pi\hbar}   \int_{-\infty}^{V^\ddagger} e^{-\tau E/\hbar+nW_0(E)/\hbar}  \bar{Z}_{n}(E;\tau) \left[1+\mathcal{O}(\hbar) \right]  \mathrm{d}E \label{eq:J_n_unintegrated_summed}
\end{equation}
where now we have a slightly different effective ``partition function''
\begin{equation}
    \bar{Z}_{n}(E;\tau) =  \sum_{\mathbf{n}} e^{-\tau{V}_{\mathbf{n},1}+n\left[W_{\mathbf{n},1}(E)+W_0'(E)V_{\mathbf{n},1}\right]}\label{eq:generalised_partition_function_2}.
\end{equation}
Asymptotic evaluation of this integral then gives
\begin{equation}
    J_{n,0,\mathrm{s}}(\hbar) = \frac{1}{2\pi}e^{-\tau V^\ddagger/\hbar} \frac{1}{n\tau_{\rm c} + \tau } Z_{ h}^\ddagger(\tau).
\end{equation}

The sums over $J_n$ and $I_n$ can then be combined together to give
\begin{equation}
    \begin{aligned}
        \!k Z_r&\sim \sum_{n=-\infty}^\infty (-1)^{n} \frac{1}{2\pi} e^{-\tau {V}^\ddagger/\hbar} \frac{1}{n{\tau}_{\rm c}+\tau} Z_h^{\ddagger}(\tau),
    \end{aligned}
\end{equation}
and the sum over $n$ can be evaluated using the identity
\begin{equation}
    \sum_{n=-\infty}^\infty \frac{(-1)^n}{\tau+n{\tau}_{\rm c}} =\frac{{\omega}}{2\sin({\omega}\tau/2)}.
\end{equation}
We therefore arrive at the leading order term in the asymptotic expansion of the rate above the crossover temperature $\tau<\tau_{\rm c}$ as
\begin{equation}
    kZ_r\sim  \frac{1}{2\pi} e^{-\tau {V}^\ddagger/\hbar} \frac{{\omega}}{2\sin({\omega}\tau/2)} Z_h^{\ddagger}(\tau) \equiv k_{\text{m-pb}}Z_r
\end{equation}
which is simply the well known multidimensional parabolic barrier approximation to the rate.

\subsubsection{First-order correction}
 Making use of the standard asymptotic result from Appendix~\ref{app:steepest_descent_integration} the first-order correction to $I_{n,0,\rm s}(\hbar)$ is given by 
 \begin{equation}
     \alpha_{n,1,\mathrm{s}} = \frac{Z_n'(V^\ddagger;\tau)}{Z_n(V^\ddagger;\tau)}\frac{1}{\tau-n\tau_{\rm c}}-\frac{nW''_0(V^\ddagger)}{(\tau-n\tau_{\rm c})^2}-\left\langle \mu_{\mathbf{n},n}(V^\ddagger)\right\rangle.
 \end{equation}
Using the fact that $\tilde{W}_{\mathbf{n}}(\tilde{V}_{\mathbf{n}})=0$ at all orders in $\hbar$ one can show
 \begin{equation}
     W_{\mathbf{n},2}(V^\ddagger)=-W_0'(V^\ddagger)V_{\mathbf{n},2}-\frac{1}{2}W_0''(V^\ddagger)V_{\mathbf{n},1}^2-W_{\mathbf{n},1}'(V^\ddagger)V_{\mathbf{n},1}
 \end{equation}
 which allows us to simplify
 \begin{equation}
     \langle \mu_{\mathbf{n},n}(V^\ddagger;\tau)\rangle = \tau \langle V_{\mathbf{n},2} \rangle.
 \end{equation}
Combining this with the identity
\begin{equation}
\begin{aligned}
    \frac{Z'_n(V^\ddagger;\tau)}{Z_n(V^\ddagger;\tau)} &= \frac{\sum_{\mathbf{n}} -n\left[W_{\mathbf{n},1}'(V^\ddagger)+W_0''(V^\ddagger)V_{\mathbf{n},1}\right] e^{-\tau{V}_{\mathbf{n},1}}}{\sum_{\mathbf{n}}  e^{-\tau{V}_{\mathbf{n},1}}} \\
    &= -n[\langle W_{\mathbf{n},1}'(V^\ddagger)\rangle+W_0''(V^\ddagger)\langle {V}_{\mathbf{n},1} \rangle]
    \end{aligned}
\end{equation}
we can write
\begin{equation}
     \alpha_{n,1,\mathrm{s}} = \frac{-n[\langle W_{\mathbf{n},1}'(V^\ddagger)\rangle+W_0''(V^\ddagger)\langle {V}_{\mathbf{n},1} \rangle]}{\tau-n\tau_{\rm c}}-\frac{nW''_0(V^\ddagger)}{(\tau-n\tau_{\rm c})^2}- \tau \langle V_{\mathbf{n},2} \rangle.
 \end{equation}

Following the same logic the first-order correction to $J_{n,0,\rm s}$ is given by
\begin{equation}
    \bar{\alpha}_{n,1,\mathrm{s}} = \frac{n[\langle W_{\mathbf{n},1}'(V^\ddagger)\rangle+W_0''(V^\ddagger)\langle {V}_{\mathbf{n},1} \rangle]}{\tau+n\tau_{\rm c}}+\frac{nW''_0(V^\ddagger)}{(\tau+n\tau_{\rm c})^2}-\tau \langle V_{\mathbf{n},2} \rangle.
\end{equation}
Hence, combining the two sums we have that to first order the rate above crossover is given by
\begin{equation}
    \begin{aligned}
        \!k Z_r&\sim \sum_{n=-\infty}^\infty (-1)^{n} \frac{1}{2\pi} e^{-\tau {V}^\ddagger/\hbar} \frac{1}{n{\tau}_{\rm c}+\tau} Z_h^{\ddagger}(\tau)\left(1+\bar{\alpha}_{n,1,\rm s} \hbar\right).
    \end{aligned}
\end{equation}
We can again perform the sums over $n$ making use of the identities
\begin{equation}
    \sum_{n=-\infty}^\infty \frac{n(-1)^n}{(\tau+n{\tau}_{\rm c})^2} = - \frac{\omega}{2\sin(\omega\tau/2)} \frac{\omega}{2\pi}  \left(\frac{\tau  \omega/2}{\tan(\tau  \omega /2)}-1\right) 
\end{equation}
and
\begin{equation}
    \sum_{n=-\infty}^\infty \frac{n(-1)^n}{(\tau+n{\tau}_{\rm c})^3} = -\frac{{\omega}}{2\sin(
    {\omega}\tau/2)} G(\tau,{\omega}) 
\end{equation}
where $G(\tau,\omega)$ is given in Eq.~(\ref{eq:G_tau_omega}).
Hence, combining these results together we arrive at Eq.~(\ref{eq:1st_order_sphaleron}) given in the main text.

\subsection{Asymptotic Evaluation of One-Dimensional Integrals}\renewcommand{\theequation}{I\arabic{equation}}
\setcounter{equation}{0}
\label{app:steepest_descent_integration}
Consider an integral of the form
\begin{equation}
    I(\lambda) = \int_{x_{0}}^\infty g(x) e^{-f(x)/\lambda} \mathrm{d}x. 
\end{equation}
If $f(x)$ has a global minimum, $f'(x^\star)=0$, then there are three possibilities: (A) The stationary point is inside the integration range, $x^\star>x_0$; (B) The stationary point is outside of the integration range, $x^\star<x_0$; (C) The stationary point is at the boundary, $x_0=x^\star$. Each of these three cases result in different asymptotic behaviours as $\lambda\to0$, and hence give rise to different asymptotic series. In principle one can combine these series into a single uniform asymptotic series that is valid in all three cases. However, for the present purpose we simply recap the standard asymptotic series for cases (A) and (B). See Ref.~\citenum{BenderBook} for a more detailed discussion.

Considering first case (A), where $x^\star>x_0$, one can derive the asymptotic behaviour by expanding the exponent and the pre-exponential about the stationary point
and then making the variable transformation $u=(x-x^\star)/\sqrt{\lambda}$. In the limit that $\lambda\to0$ the lower limit of integration also goes to infinity and at leading order we are left with
\begin{equation}
\begin{aligned}
    I(\lambda)\sim I^{\rm A}_0(\lambda) &= e^{-f(x^\star)/\lambda} \sqrt{\lambda} \int_{-\infty}^\infty g(x^\star)e^{-f''(x^\star)u^2/2}  \mathrm{d}u \\
    & = e^{-f(x^\star)/\lambda} \sqrt{\frac{2\pi\lambda}{f''(x^\star)}}g(x^\star).
    \end{aligned}
\end{equation}
Expanding to first order then gives 
\begin{equation}
\begin{aligned}
    I(\lambda) \sim I^{\rm A}_0(\lambda)\Big(1+\lambda a_1^{\rm A} + \mathcal{O}(\lambda^2)\Big)
    \end{aligned}
\end{equation}
with
\begin{equation}
\begin{aligned}
    a_1^{\rm A}=&\frac{g^{(2)}(x^\star)}{2g(x^\star)f^{(2)}(x^\star)}-\frac{g^{(1)}(x^\star)f^{(3)}(x^\star)}{2g(x^\star)[f^{(2)}(x^\star)]^{2}}\\ &-\frac{f^{(4)}(x^\star)}{8[f^{(2)}(x^\star)]^{2}}+\frac{5[f^{(3)}(x^\star)]^2}{24[f^{(2)}(x^\star)]^{3}}
    .
    \end{aligned}
\end{equation}

Consider now case (B), where $x^\star<x_0$, expanding both the pre-exponential and exponent about the boundary (which is the maximum within the integration region), and making the variable transformation $u=(x-x_0)/\lambda$, we find that the leading order term in the asymptotic series is given by 
\begin{equation}
        I(\lambda) \sim I^{\rm B}_0(\lambda) = e^{-f(x_0)/\lambda}  g(x_0) \frac{\lambda}{f'(x_0)}.
\end{equation}
Expanding to first order then gives
\begin{equation}
\begin{aligned}
    I(\lambda) \sim I^{\rm B}_0(\lambda)\Big(1+\lambda a_1^{\rm B} + \mathcal{O}(\lambda^2)\Big)
    \end{aligned}
\end{equation}
with
\begin{equation}
        a_1^{\rm B} = \frac{g'(x_0)}{g(x_0)}\frac{1}{f'(x_0)}-\frac{ f''(x_0)}{[f'(x_0)]^2}.
\end{equation}

Note that, for integrals where the finite limit occurs at the upper boundary of the form 
\begin{equation}
    J(\lambda) = \int_{-\infty}^{x_0} g(x) e^{-f(x)/\lambda} \mathrm{d}x
\end{equation}
the formulas for case (A) remain unchanged, but for case (B) the leading order expression changes to  
\begin{equation}
        J(\lambda) \sim J_0^{\rm B}(\lambda)= e^{-f(x_0)/\lambda}  g(x_0) \frac{-\lambda}{f'(x_0)}
\end{equation}
and the expression for the first-order correction remains unchanged.

\bibliography{references,extra_refs}

\end{document}